\documentclass[11pt,aps,preprint,nofootinbib,superscriptaddress]{revtex4}
\usepackage[active]{srcltx}
\usepackage[utf8]{inputenc}
\usepackage{latexsym}
\usepackage{amsmath}
\usepackage{times,txfonts}
\usepackage{graphicx,color}
\usepackage[colorlinks=true,linkcolor=blue,urlcolor=blue,citecolor=blue,pdfusetitle]{hyperref}

\begin{document}

\title{Supersymmetric Quantum Spherical Spins with Short-Range Interactions}

\author{L. V. T. Tavares }
\email{ltavares@uel.br}
\affiliation{Departamento de F\'isica, Universidade Estadual de Londrina, \\
Caixa Postal 10011, 86057-970, Londrina, PR, Brasil}

\author{L. G. dos Santos}
\email{lgsantos@uel.br}
\affiliation{Departamento de F\'isica, Universidade Estadual de Londrina, \\
Caixa Postal 10011, 86057-970, Londrina, PR, Brasil}

\author{G. T. Landi}
\email{gtlandi@if.usp.br}
\affiliation{Instituto de F\'isica da Universidade de S\~ao Paulo,  05314-970 S\~ao Paulo, Brazil}

\author{Pedro R. S. Gomes}
\email{pedrogomes@uel.br}
\affiliation{Departamento de F\'isica, Universidade Estadual de Londrina, \\
Caixa Postal 10011, 86057-970, Londrina, PR, Brasil}

\author{P. F. Bienzobaz}
\email{paulabienzobaz@uel.br}
\affiliation{Departamento de F\'isica, Universidade Estadual de Londrina, \\
Caixa Postal 10011, 86057-970, Londrina, PR, Brasil}


\begin{abstract}

This work is dedicated to the study of a supersymmetric quantum spherical spin system with short-range interactions. We examine the critical properties both a zero and finite temperature. The model undergoes a quantum phase transition at zero temperature without breaking supersymmetry. At finite temperature the supersymmetry is broken and the system exhibits a thermal phase transition. We determine the critical dimensions and compute critical exponents. In particular, we find that the model is characterized by a dynamical critical exponent $z=2$. We also investigate properties of correlations in the one-dimensional lattice. Finally, we explore the connection with a nonrelativistic version of the supersymmetric $O(N)$ nonlinear sigma model and show that it is equivalent to the system of spherical spins in the large $N$ limit.

\end{abstract}
\maketitle


\section{Introduction}\label{S1}

This work is dedicated to the study of a supersymmetric quantum spherical spin with short-range interactions. This model refers to a supersymmetric extension of a system of quantum spherical spins, i.e., a lattice model involving continuous spin variables, $-\infty<S_{\bf r}<\infty$, attached to each site of a hypercubic lattice and subject to the spherical constraint $\sum_{\bf r}S_{\bf r}^2=N$, where $N$ is the total number of sites of the lattice. The classical Hamiltonian is \cite{Berlin}, 
\begin{equation}
\mathcal{H}_c=\frac{1}{2} \sum_{{\bf r},{\bf r}'}J_{{\bf r},{\bf r}'}S_{\bf r}S_{{\bf r}'}.
\end{equation}
A quantum version can be constructed by introducing a nontrivial dynamics to the spins by means of a kinetic term involving the conjugated momentum of $S_{\bf r}$, for example, of the form $\sim \sum_{\bf r} P_{\bf r}^2$ \cite{Obermair}. Thus canonical quantization can be immediately carried out. Alternatively, quantum features may be introduced using  path integrals \cite{Gracia}. Classical and quantum versions of spherical spins have been intensively studied in a number of situations \cite{Joyce,Obermair,Henkel,Nieuwenhuizen,Vojta1,Gracia}. This is mainly due to the fact that they are exactly soluble, 
even in the presence of an external field and may present non-mean-field critical exponents. 
The spherical model thus constitutes a fruitful framework to examine a number of questions of great interest in the study of critical phenomena and phase transitions.

The supersymmetric model has its origin in the search for further generalizations of the previous studies while keeping the same spirit of the spherical model, i.e., without loosing the remarkable properties mentioned above. Usually the spherical versions involve only scalar spin variables like $S_{\bf r}$ at each site, that after quantization correspond to bosonic degrees of freedom. A natural generalization of this situation consists in adding degrees of freedom of fermionic character at each site. This can be done in a controllable way by requiring that the new degrees of freedom enter on an equal footing with the bosonic ones, i.e., by requiring that the whole system be supersymmetric. 

Over the years, the arena of supersymmetry has become wider, far beyond its original conception in describing elementary particles and strings \cite{Dine}. The efforts to place supersymmetry in a broader context, outside high-energy physics, is largely due to the unfortunate dissonance with other physical theories, where in general the elegance of theoretical descriptions is graced by experimental evidences. After all, it would be rather disappointing if nature, at any level, does not choose to enjoy such a remarkable symmetry. Apart from the high-energy scenario, favorable places to find supersymmetry are in systems involving many distinct degrees of freedom, like the ones frequently considered in statistical mechanics and condensed matter physics. One of the first examples is the supersymmetry arising in the tricritical Ising model \cite{Friedan1,Qiu}. More recently, a number of studies has reported that supersymmetry does emerge in special points of the parameter space in several quantum models \cite{Lee,Affleck,Ponte,Grover,Yao,Jian}. Of course, in all these cases, the supersymmetry is thought in an effective sense, different from the original purpose. It is in the effective sense that we treat the supersymmetric model discussed in this work, i.e., as describing effective quantum degrees of freedom. We remember that the quantum spherical model is an akin of the quantum rotors \cite{Vojta1}, which in turn can be used to describe low-energy excitations of many systems \cite{Sachdev}. Therefore, it is expected that its supersymmetric counterpart can also play an interesting role in such systems.

A proper way to obtain the supersymmetric model is by proceeding with the superspace formalism \cite{Gates}. This is so because we have to generalize the spherical constraint in compliance with supersymmetry and working in the superspace takes this into account automatically. The minimal supersymmetric model requires extended supersymmetry with $\mathcal{N}=2$ supercharges. In this case, the usual spin variable $S_{\bf r}$ is replaced by the superfield $\Phi_{\bf r}(t,\theta,\bar{\theta})=S_{\bf r}+\bar\theta \psi_{\bf r}+\bar{\psi}_{\bf r}\theta+\bar{\theta}\theta F_{\bf r}$, which contains, in addition to the usual spin variable $S_{\bf r}$, two fermionic degrees of freedom, $\psi_{\bf r}$ and $\bar{\psi}_{\bf r}$, and an auxiliary (nonphysical) bosonic degree of freedom, $F_{\bf r}$. The Grassmann variables $\theta$ and $\bar{\theta}$, together with the time, $t$, are the coordinates of the superspace. The generalization of the spherical constraint corresponds simply to 
\begin{equation}
\sum_{\bf r}\Phi_{\bf r}^2=N.
\label{0.1a}
\end{equation} 
With these ingredients, we can write the action in the superspace, 
\begin{equation}
S=\int dt d\theta d\bar{\theta} \left[\frac{1}{2}\sum_{\bf r} \bar{D}\Phi_{\bf r} D\Phi_{\bf r} 
+\frac{1}{2}\sum_{{\bf r},{\bf r}'}U_{{\bf r},{\bf r}'}\Phi_{\bf r}\Phi_{{\bf r}'} 
-\Xi\left(\sum_{\bf r}\Phi_{\bf r}^2-N\right)\right],
\label{0.1}
\end{equation}
where $D$ and $\bar{D}$ are the supercovariant derivatives, 
\begin{equation}
D\equiv -\frac{\partial}{\partial\bar\theta}+i\theta\frac{\partial}{\partial t}~~~
\text{and}~~~\bar{D}\equiv \frac{\partial}{\partial\theta}-i\bar\theta\frac{\partial}{\partial t},
\label{0.2}
\end{equation}
and 
\begin{equation}
\Xi(t,\theta,\bar{\theta})= \gamma+\bar{\theta}\xi+\bar{\xi}\theta+\bar{\theta}\theta \mu,
\label{0.3}
\end{equation}
is a  Lagrange multiplier that enforces the constraints
\begin{equation}
\sum_{\bf r}S_{\bf r}^2=N,~~~\sum_{\textbf{r}}S_{\textbf{r}}\psi_{\textbf{r}} =0,~~~\sum_{\textbf{r}}S_{\textbf{r}}\bar{\psi}_{\textbf{r}} =0,~~~\text{and}~~~\sum_{\textbf{r}}S_{\textbf{r}}F_{\textbf{r}}=\sum_{\bf r}\bar{\psi}_{\textbf{r}}\psi_{\textbf{r}}.
\label{0.4}
\end{equation}
The interaction energy $U_{{\bf r},{\bf r}'}=U(|{\bf r}-{\bf r}'|)$ entering the action (\ref{0.1}) controls the range of the interaction. 

It is also instructive to write the Lagrangian in terms of components. After integrating over the Grassmann variables, we find
\begin{eqnarray}
L& =&\frac{1}{2}\sum_{\bf r}\dot{S}_{\bf r}^2
+\frac{1}{2}\sum_{\bf r}F_{\bf r}^2+i\sum_{\bf r}\bar{\psi}_{\bf r}\dot{\psi_{\bf r}}
+\sum_{{\bf r},{\bf r}'}U_{{\bf r},{\bf r}'}\left(
S_{\bf r}F_{{\bf r}'} -\bar{\psi}_{\bf r}\psi_{{\bf r}'}\right)\nonumber\\
&+&\gamma\sum_{\bf r}\left(F_{\bf r}S_{\bf r}-\bar{\psi}_{\bf r}\psi_{\bf r}\right)
-\sum_{\bf r}\left(\bar{\psi}_{\bf r}\xi +\bar{\xi}\psi_{\bf r} \right)S_{\bf r}
-\mu\left(\sum_{\bf r}S_{\bf r}^2-N\right),
\label{2.19a}
\end{eqnarray}
up to redefinitions of the Lagrange multipliers to absorb unimportant numerical factors. The corresponding supersymmetry transformations that leave this Lagrangian invariant are
\begin{equation}
\epsilon:~~~\delta_{\epsilon}S_{\bf r}=\bar{\psi}_{\bf r}\epsilon,~~~
\delta_{\epsilon}\psi_{\bf r}=-i\dot{S}_{\bf r}\epsilon+
F_{\bf r}\epsilon,~~~
\delta_{\epsilon}\bar{\psi}_{\bf r}=0,~~~\text{and}~~~\delta_{\epsilon}F_{\bf r}=i\dot{\bar\psi}_{\bf r}\epsilon;
\label{2.11a}
\end{equation}
and
\begin{equation}
\bar{\epsilon}:~~~\delta_{\bar\epsilon}S_{\bf r}=\bar\epsilon{\psi}_{\bf r},~~~
\delta_{\bar\epsilon}\psi_{\bf r}=0,~~~
\delta_{\bar\epsilon}\bar{\psi}_{\bf r}=i\dot{S}_{\bf r}\bar\epsilon+
F_{\bf r}\bar\epsilon,~~~\text{and}~~~\delta_{\bar\epsilon}F_{\bf r}=-i\bar\epsilon\dot{\psi}_{\bf r},
\label{2.12a}
\end{equation}
where $\epsilon$ and $\bar{\epsilon}$ are the parameters (Grassmann) of the transformation. 

In Ref.~\cite{Lucas} we have provided a detailed discussion on the construction of the supersymmetric model as well as a comparison with the previous studies. In Ref.~\cite{Lucas}  we also present the calculation of the partition function via the saddle point method for arbitrary interactions depending only on the distance between the sites, followed by an extensive analysis of the mean-field critical behavior (obtained by setting $U_{{\bf r},{\bf r}'}\rightarrow U/N$). 
Such analysis provides a qualitative understanding of the general pattern of phase transitions in the system. 

The present work is a direct continuation of Ref.~\cite{Lucas}. 
Here we move beyond the mean-field analysis and examine in detail the  critical properties for the more interesting case of short-range interactions, 
\begin{eqnarray}
U_{\textbf{r},\textbf{r}'} \equiv U\sum_{i=1}^{d}\left(\delta_{\textbf{r},\textbf{r}'+\mathbf{e}^{i}}+\delta_{\textbf{r},\mathbf{r}'-\mathbf{e}^{i}}\right),
\label{0.5}
\end{eqnarray}
where $U$ is the interaction energy that can be positive (ferro) or negative (anti-ferro), and we are considering a $d$-dimensional hypercubic lattice with $\mathbf{e}^{i}$ being a set of orthogonal unit vectors,
\begin{eqnarray}
\left\{ \mathbf{e}^{i}\right\}  & = & \left\{ \left(1,0,\ldots,0\right);\left(0,1,0,\ldots,0\right);\ldots;\left(0,\ldots,0,1\right)\right\}.
\label{0.6}
\end{eqnarray}
By studying the convergence properties of the saddle point equations coming from the constraints, we determine the critical dimensions of the model in both cases of zero and finite temperature, and also compute the critical exponents of magnetization and susceptibility.  The analysis of the solutions of the saddle point equations shows that, in the case of zero temperature, there is no spontaneous supersymmetry breaking, such that a quantum phase transition takes place with supersymmetry preserved. In the case of finite temperature, on the other hand, supersymmetry is broken by thermal effects and there is an additional solution of saddle point equations, which in turn changes the critical exponent of the susceptibility, and hence the universality class of the phase transition. By comparing the shift in the critical dimensions in the cases of zero and finite temperature we extract a dynamical critical exponent $z=2$. This follows from the usual classical-quantum mapping connecting thermal critical phenomena in $D$ spatial dimensions and quantum critical phenomena in the reduced $d=D-z$ spatial dimensions. In this case, $z$ parametrizes the difference between the behavior of the correlation length, $\xi$, and the correlation time, $\tau_c$, near the critical point, according to $\xi\sim|t|^{-\nu}$ and $\tau_c=\xi^{z}\sim|t|^{-z\nu}$, where $t$ measures the distance from the critical point \cite{MVojta,Sachdev}. \footnote{If we interpret the time as an additional spatial coordinate, we can make contact with the anisotropy exponent $\theta$ usually defined in magnetic systems involving competing interactions (exhibiting a Lifshitz point), which is given by $\theta=\frac{\nu_{\|}}{\nu_{\bot}}$ \cite{Henkel2,Shpot}. In this relation, $\nu_{\bot}$ is associated with the correlation length in the directions with only first neighbor interactions, and $\nu_{\|}$ with the correlation length along the directions with the competing interactions, i.e., $\xi_{\bot}\sim |t|^{-\nu_{\bot}}$ and $\xi_{\|}\sim |t|^{-\nu_{\|}}$. Therefore, we have the identification $\theta \Leftrightarrow\frac{1}{z}$.}

Another remarkable property of classical and quantum spherical models is the connection with nonlinear sigma-type models in the limit of large number of fields. Some of the specific relations are 
\begin{eqnarray}
\text{Classical Spherical Model} ~~&\Longleftrightarrow&~~ \text{Classical Heisenberg Model~\cite{Stanley}}\nonumber\\ 
\text{Quantum Spherical Model}~~&\Longleftrightarrow&~~O(N)~\text{Nonlinear Sigma Model~\cite{Vojta1,Gomes2}}\nonumber\\
\text{Gauged Quantum Spherical Model}~~&\Longleftrightarrow&~~CP^{(N)}~ \text{Model~\cite{Bienzobaz3}}.
\label{9.0}
\end{eqnarray}
This is an appealing property since it softens the issue with the long-range interactions effectively introduced by the spherical constraint, relating such models with ones which involve exclusively short-range interactions.

In this context, we address the question of whether the supersymmetric extension considered here has an equivalent description in terms of some field theoretical model in the large $N$ limit. We shall see that it has also a counterpart version given in terms of a supersymmetric nonlinear sigma model, enlarging the set of  equivalences. One important guide in this direction is the dynamical critical exponent $z=2$, which implies an anisotropic scaling between the correlation length and the correlation time. This feature leads us to look for continuum field theories that embody this and hence should not be Lorentz invariant. They are referred to as Lifshitz field theories, since this type of nonrelativistic theory is used in the description of quantum Lifshitz points \cite{Ardonne}.

In the context of high-energy physics, field theories of Lifshitz type have attracted a lot of attention in the recent years, mainly due to the possibility of being implemented in quantum gravity \cite{Horava}. The anisotropic scaling characterized by $z=2$ amounts to the introduction of higher spatial derivative operators in the action, which in effect improves the UV behavior without breaking unitarity, rendering the theory power-counting renormalizable in four spacetime dimensions. In this setting, the local Lorentz invariance is expected to emerge in the low-energy sector. Since this proposal, many studies have been conducted to examine renormalization group flows in such theories, with the general pattern pointing out that the restoration of Lorentz invariance depends in general on fine-tunings \cite{Iengo,Gomes8}.

This work is organized as follows. In Sec. \ref{S2} we review the computation of the partition function via saddle point method and discuss the corresponding solutions according to supersymmetry breaking. Sec. \ref{S3} is dedicated to the study of the quantum critical behavior for the case of short-range interactions, including the computation of some critical exponents. In Sec. \ref{S4} the previous analysis is extended to the case of finite temperatures. In Sec. \ref{S5} we compute bosonic correlations functions in the one-dimensional case, aiming to investigate possible oscillations due to competing interactions. In Sec. \ref{S6} we examine the connection with the nonrelativistic supersymmetric nonlinear sigma model. We conclude in Sec. \ref{S7} with a summary and additional comments.

\section{Partition Function and Supersymmetry Breaking}\label{S2} 

In this section we briefly review the saddle point computation of the partition function \cite{Lucas}. In terms of  imaginary time, $t=-i\tau$, with $\tau\in [0,\beta]$ and $\beta=1/T$,  the partition function reads,
\begin{eqnarray}
\mathcal{Z} & = &\int\mathcal{D}\Omega\exp\left\{
-\int_0^\beta d\tau\left[ L_E+H_B\sum_{\bf r}S_{\bf r}
+H_F\sum_{\bf r}\bar{\psi}_{\bf r}\psi_{\bf r}\right]\right\},
\label{eq:Z1}
\end{eqnarray}
where the measure $\mathcal{D}\Omega$ corresponds to the integral over all fields as well as over the Lagrange multipliers that implement the supersymmetric constraints,
$\mathcal{D}\Omega\equiv \mathcal{D}S\mathcal{D}F\mathcal{D}\psi\mathcal{D}\bar{\psi}\mathcal{D}\mu\mathcal{D}\gamma\mathcal{D}\xi\mathcal{D}\bar{\xi}$, and $L_E$ is the Euclidean version of the Lagrangian in (\ref{2.19a}),
\begin{eqnarray}
L_E&=&\frac{1}{2g}\sum_{\bf r}\left(\frac{\partial{S}_{\bf r}}{\partial\tau}\right)^2
-\frac{1}{2}\sum_{\bf r}F_{\bf r}^2+\frac{1}{\sqrt{g}}\sum_{\bf r}\bar{\psi}_{\bf r}\frac{\partial \psi_{\bf r}}{\partial\tau}
-\sum_{{\bf r},{\bf r}'}U_{{\bf r},{\bf r}'}\left(
S_{\bf r}F_{{\bf r}'} -\bar{\psi}_{\bf r}\psi_{{\bf r}'}\right)\nonumber \\
&-&\gamma\left(\sum_{\bf r}F_{\bf r}S_{\bf r}-\sum_{\bf r}\bar{\psi}_{\bf r}\psi_{\bf r}\right)
+\sum_{\bf r}\bar{\psi}_{\bf r}\xi S_{\bf r}
+\sum_{\bf r}\bar{\xi}\psi_{\bf r} S_{\bf r}
+\mu\left(\sum_{\bf r}S_{\bf r}^2-N\right).
\label{2.19ab}
\end{eqnarray}
Notice that we have introduced a parameter $g$ in the Euclidean Lagrangian through the rescaling $\tau\rightarrow\sqrt{g}\tau$, which measures the quantum fluctuations in the system. In the case of zero temperature, this is the parameter that controls the distance of the quantum critical point, playing a role similar to the temperature in the case of a phase transition driven by thermal fluctuations. We have also included in the partition function two external fields, $H_{B}$ and $H_{F}$, so that by taking derivatives with respect to them we obtain respectively the order parameter $\langle \sum_{\bf r}S_{\bf r}\rangle$ and the fermionic condensate $\langle \sum_{\bf r}\bar{\psi}_{\bf r}\psi_{\bf r}\rangle$.

The integral over the fields $S,\psi,\bar{\psi}$, and $F$ are at most quadratic and thus and can be directly performed. This leads to 
\begin{eqnarray}
\mathcal{Z}=\int \mathcal{D}\mu \mathcal{D}\gamma \mathcal{D}\bar{\xi}
\mathcal{D}\xi\text{e}^{-NS_{eff}},
\label{0.7}
\end{eqnarray}
with the effective action given by
\begin{eqnarray}
S_{eff}&\equiv&\frac{1}{2N}\text{Tr}\sum_{\bf q}\ln \left[
-\frac{1}{2g}\frac{\partial^2}{\partial \tau^2}+\mu+
\frac{(U({\bf q})+\gamma)^2}{2}
\right]\nonumber\\
&-&\frac{1}{N}\text{Tr}\sum_{\bf q}\ln\left[
\frac{1}{\sqrt{g}}\frac{\partial}{\partial \tau}+ U({\bf q})+\gamma
+H_F-\frac{1}{2}\xi\mathcal{O}^{-1}_{\bf q}\bar{\xi}
\right]\nonumber\\
&-&\frac{1}{4}\int_0^{\beta}d\tau\frac{H_B^2}{\mu+\frac{\left[U(0)+\gamma\right]^2}{2}}
-\int_0^{\beta}d\tau\mu,
\label{0.8}
\end{eqnarray}
where $U({\bf q})$ is the Fourier transform of the interaction $U_{{\bf r},{\bf r}'}\equiv U(|{\bf r - r'}|)$, 
\begin{eqnarray}
{U}({\bf q})=\sum_{\bf r-r'}U(|{\bf r-r'}|)
\text{e}^{i{\bf q}\cdot{\bf h}},
\label{0.9}
\end{eqnarray}
and the operator $\mathcal{O}_{\bf q}$ is defined as
\begin{eqnarray}
\mathcal{O}_{\bf q}\equiv-\frac{1}{2g}\frac{\partial^2}{\partial\tau^2}+\mu+\frac{1}{2}\left[U({\bf q})+\gamma\right]^2.
\label{0.10}
\end{eqnarray}

The remaining integrals in (\ref{0.7}) can be evaluated through the saddle point method, which becomes exact in the thermodynamic limit $N\rightarrow\infty$. The saddle point equations are determined by the conditions
\begin{eqnarray}
\frac{\delta S_{eff}}{\delta \mu}
=\frac{\delta S_{eff}}{\delta \gamma}
=\frac{\delta S_{eff}}{\delta \xi}
=\frac{\delta S_{eff}}{\delta \bar{\xi}}=0.
\label{0.11}
\end{eqnarray}
The last two equations (the fermionic ones) are trivially satisfied with $\xi=\bar{\xi}=0$, whereas the bosonic ones yield to the constraint equations for the parameters $\mu$ and $\gamma$,
\begin{eqnarray}
1=\frac{H_B^2}{4\left[\mu+\frac{1}{2}(U(0)+\gamma)^2\right]^2}
+\frac{1}{2N}\sum_{\bf q}\frac{g}{w_{\bf q}^B}\coth\left(
\frac{\beta}{2}w_{\bf q}^B
\right),
\label{0.12}
\end{eqnarray}
and
\begin{eqnarray}
0&=&\frac{H_B^2}{4\left[\mu+\frac{(U(0)+\gamma)^2}{2}\right]^2}~
\left[U(0)+\gamma\right]
+\frac{1}{2N}\sum_{\bf q}\frac{g}{w_{\bf q}^B}~
[U({\bf q})+\gamma]\coth\left(
\frac{\beta}{2}w_{\bf q}^B
\right)\nonumber\\
&-&\frac{1}{2N}\sum_{\bf q}\frac{g}{w_{\bf q}^F}[U({\bf q})+\gamma+H_F]
\tanh\left(\frac{\beta}{2}w_{\bf q}^F\right),
\label{0.13}
\end{eqnarray}
with the bosonic and fermionic frequencies defined as
\begin{eqnarray}
\left(w_{\bf q}^B\right)^2\equiv 2g\left\{\mu+\frac{1}{2}\left[U({\bf q})+\gamma\right]^2\right\}
~~~~\text{and}~~~~(w_{\bf q}^F)^2= g\left[U({\bf q})+\gamma+H_F\right]^2.
\label{0.14}
\end{eqnarray}
The analysis of convergence properties of equations (\ref{0.12}) and (\ref{0.13}) determine the critical behavior of the model. In the next section, we will perform a detailed analysis  for the case of short-range interactions (\ref{0.5}), whose Fourier transform takes the form,
\begin{eqnarray}
U({\bf q})= 2 U\sum_{i=1}^d\cos q_i.
\label{0.15}
\end{eqnarray}

Before doing so, however, it is instructive to look at the free energy of the system, $f=\frac{1}{\beta} S_{eff}$, which gives 
\begin{eqnarray}
f&=&-\frac{H_B^2}{4\left[\mu+\frac{1}{2}\left(U(0)+\gamma\right)^2\right]}-\mu
+\frac{1}{\beta N}\sum_{\bf q}\ln\left[\frac{
2\sinh\left(\frac{\beta}{2}w_{\bf q}^B\right)}
{2\cosh\left(
\frac{\beta}{2}w_{\bf q}^F\right)}\right].
\label{free_energy}
\end{eqnarray}
In the limit $T\to 0$ this expression reduces to the ground state energy, that in the absence of the external fields reads, 
\begin{eqnarray}
\frac{E_0}{N}&=&-\mu+\frac{1}{2N}\sum_{\bf q} (w_{\bf q}^B-w_{\bf q}^F)\nonumber\\
&=&-\mu+\frac{1}{2N}\sum_{\bf q} 
\left\{
\left[2g\left(\mu+\frac{1}{2}\left(U({\bf q})+\gamma\right)^2\right)\right]^{\frac{1}{2}}
-\left[2g\left(\frac{1}{2}\left(U({\bf q})+\gamma\right)^2\right)\right]^{\frac{1}{2}}
\right\}.
\label{0.16}
\end{eqnarray}
We see that it vanishes only for $\mu=0$, independently of $\gamma$. As a nonvanishing ground state energy is a diagnosis of supersymmetry breaking this implies that any solution of (\ref{0.12}) and (\ref{0.13}) with $\mu\neq 0$ corresponds to a spontaneous supersymmetry breaking. In the case of finite temperature, supersymmetry is always broken by thermal effects, independent of the values taken by $\mu$ and $\gamma$.

\section{Quantum Critical Behavior}\label{S3}

\subsection{Behavior of the Lagrange multiplier $\gamma$ near criticality}

To study the quatum critical behavior we have to analyze the spherical constraints (\ref{0.12}) and (\ref{0.13}) in the limit of zero temperature ($\beta\rightarrow\infty$), which enables us to obtain the parameters $\mu$ and $\gamma$ as a function of $g$, $H_B$, and $H_F$. It is helpful to recall here that the parameter $\mu$ implements the usual constraint $\sum_{\bf r}S_{\bf r}^2=N$, whereas $\gamma$ implements the constraint $\sum_{\bf r}\left(F_{\bf r}S_{\bf r}-\bar{\psi}_{\bf r}\psi_{\bf r}\right)=0$. Thus $\gamma$ is responsible for the coupling between bosonic and fermionic degrees of freedom.

By considering firstly $H_B=H_F=0$, it is immediate to verify that the expression (\ref{0.13}) is satisfied only if $\mu=0$  independent of the value of $\gamma$, 
implying that supersymmetry is not spontaneously broken in this model. In the thermodynamic limit, 
\begin{eqnarray}
\frac{1}{N}\sum_{\bf q} \rightarrow\int \frac{d^d{q}}{(2\pi)^d},
\label{5}
\end{eqnarray}
the critical behavior is then governed by Eq. (\ref{0.12}) with $H_B=\mu=0$,
\begin{eqnarray}
1=\frac{1}{N}\sum_{\bf q}\frac{g}{2w_{{\bf q}}^{B}}=
\frac{\sqrt{g}}{2}\int \frac{d^d{q}}{(2\pi)^d}\frac{1}{|\gamma+2U\sum_i \cos(q_i)|},
\label{6}
\end{eqnarray}
which involves only on the Lagrange multiplier $\gamma$ that carries the information of interaction between bosons and fermions. This is a crucial difference compared to the non-supersymmetric counterpart of the model. The model will exhibit a critical point if the momentum integral in (\ref{6}) does converge even when the denominator approaches to zero. The critical point is thus located at $\gamma_c+2U\text{max}(\sum_i\cos q_i)=0$ if $\gamma$ and $U$ have opposite signs, and at $\gamma_c+2U\text{min}(\sum_i\cos q_i)=0$ if $\gamma$ and $U$ have the same sign. By writing the on-shell version of the action (\ref{2.19a}), we can see that the product $\gamma U$ is effectively the interaction energy between first-neighbors (we show this explicitly in Sec. \ref{S5}, Eq. (\ref{ssa})), so that $\gamma U<0$ corresponds to a ferromagnetic interaction whereas $\gamma U>0$ to an anti-ferromagnetic one. For concreteness, throughout this work we consider that $\gamma$ and $U$ have opposite signs, say $\gamma>0$ and $U<0$, where
\begin{eqnarray}
\gamma_c = 2|U|d.
\label{7}
\end{eqnarray}

To proceed let us assume momentarily that $\gamma+2U\sum_i \cos(q_i)>0$, such that we can get rid of the absolute value in the denominator of (\ref{6}) (we shall see in the numerical solution that when $\gamma>0$ and $U<0$, this is indeed the case). It is convenient to rewrite (\ref{6}) with help of the identity \cite{Grad}
\begin{eqnarray}
\frac{1}{x^p}= \frac{1}{\Gamma(p)}\int_0^{\infty} dt~
t^{p-1}\exp\left(-xt\right),~~~p,x>0,
\label{8}
\end{eqnarray} 
so that the constraint equation is expressed as
\begin{eqnarray}
1&=&\frac{\sqrt{g}}{4|U|}\int_{-\pi}^{\pi}\frac{d^dq}{(2\pi)^d}\int_0^{\infty}dt~\exp{\left[-\left(
\frac{\gamma}{2|U|}-\sum_i\cos q_i\right) t\right]}\nonumber\\
&=&\frac{\sqrt{g}}{4|U|}\int_0^{\infty}dt\exp\left(-\frac{\gamma}{2|U|}t\right)
\int_{-\pi}^{\pi} \frac{d^dq}{(2\pi)^d}\exp\left(t\sum_i\cos q_i\right)\nonumber\\
&=&\frac{\sqrt{g}}{4|U|}\int_0^{\infty}dt\exp\left(-\frac{\gamma}{2|U|}t\right) \left [\text{I}_0(t)\right]^d,
\label{9}
\end{eqnarray}
where $\text{I}_{\alpha}(t)$ is the modified Bessel function of the first type. The analysis now follows a standard approach in the literature. We have to investigate the convergence properties of the integral appearing in this expression,
\begin{eqnarray}
\mathcal{I}_d(\gamma)\equiv \int_0^{\infty}dt\exp\left(-\frac{\gamma}{2|U|}t\right) \left[\text{I}_0(t)\right]^d.
\label{11}
\end{eqnarray}
To this we use the asymptotic behaviors of $\text{I}_0(t)$,
\begin{eqnarray}
\text{I}_0(t)&\sim& 1,~~~~~~~~~~~~~t \rightarrow0, \nonumber\\
\text{I}_0(t)&\sim& \frac{\text{e}^{t}}{(2\pi t)^{\frac{1}{2}}},~~~~~t\rightarrow\infty.
\label{10}
\end{eqnarray}
These behaviors show that a potential divergence of (\ref{11}) is located in the region of large values of $t$. It is convergent at the critical point for $d>2$, which determines the lower critical dimension of the model $d_l^0=2$.  
In this case the model exhibits a critical point when $\gamma$ reaches $\gamma_{c}$, with a corresponding value $g=g_c$.
To extract the dependence on $(\gamma-\gamma_c)$ according to the dimensionality in $d>2$, we consider the derivative of $\mathcal{I}_d$ with respect to $\gamma$ in the large-$t$ region,
\begin{equation}
\mathcal{I}'_d(\gamma)\sim - \int_0^{\infty} dt ~t^{-\frac{(d-2)}{2}}\exp[-t(\gamma-\gamma_c)].
\label{10b}
\end{equation}
This expression converges at the critical point for $d>4$, which determines the upper critical dimension of the model $d_u^0=4$. For $2<d<4$, we can find the leading order contribution for $\gamma\sim \gamma_c$ by evaluating the integral in (\ref{10b}):
\begin{equation}
\mathcal{I}'_d(\gamma)\sim-(\gamma-\gamma_c)^{\frac{-(4-d)}{2}}\Gamma\left(\frac{4-d}{2}\right).
\label{10c}
\end{equation}
Integrating this expression in $\gamma$ we obtain
\begin{equation}
\mathcal{I}_d(\gamma)-\mathcal{I}_d(\gamma_c)\sim - \frac{(\gamma-\gamma_c)^{\frac{(d-2)}{2}}}{\left(\frac{d-2}{2}\right)}\Gamma\left(\frac{4-d}{2}\right).
\label{10d}
\end{equation}
Since Eq. (\ref{9}) is of the form $1/\sqrt{g}\sim \mathcal{I}_d(\gamma)$, by expanding it around the critical point and using (\ref{10d}), it follows that
\begin{equation}
\tau_g\sim(\gamma-\gamma_c)^{\frac{(d-2)}{2}},~~~\text{for}~~2<d<4,
\end{equation}
where $\tau_g\equiv(\sqrt{g}-\sqrt{g_c})/\sqrt{g_c}$.

For $d=4$ we need to be a little more careful with (\ref{11}). We also consider its derivative with respect to $\gamma$, 
\begin{equation}
\mathcal{I}'_d(\gamma)\sim -\int_0^{\infty}dt\exp\left(-\frac{\gamma}{2|U|}t\right) t \left [\text{I}_0(t)\right]^4.
\label{10e}
\end{equation}
We then split the integration region as $\int_0^{\infty}=\int_0^1+\int_{1}^{\infty}$. The integral in the first part is clearly finite and for the second part we use the asymptotic behavior in (\ref{10}),
\begin{eqnarray}
\mathcal{I}'_d(\gamma)&\sim&-\int_1^{\infty}dt\exp\left(-\frac{t}{2|U|}(\gamma-\gamma_c)\right) t^{-1} \nonumber\\
&\sim&- \Gamma\left(0,\frac{1}{2|U|}(\gamma-\gamma_c)\right),
\label{10f}
\end{eqnarray}
where $\Gamma\left(0,\frac{(\gamma-\gamma_c)}{2|U|}\right)$ is the incomplete gamma function \cite{Grad}. Its behavior for small $(\gamma-\gamma_c)$ is
\begin{equation}
\Gamma\left(0,\frac{1}{2|U|}(\gamma-\gamma_c)\right)=-\text{ln}\left[\frac{(\gamma-\gamma_c)}{2|U|}\right]+\Gamma'(1)+O(\gamma-\gamma_c),
\label{10g}
\end{equation}
where $-\Gamma'(1)$ is the Euler constant. Using this in (\ref{10f}) and integrating in $\gamma$, we get
\begin{equation}
\mathcal{I}_d(\gamma)-\mathcal{I}_d(\gamma_c)\sim (\gamma-\gamma_c) \text{ln} (\gamma-\gamma_c),
\end{equation}
that, together with Eq. (\ref{9}), implies
\begin{equation}
\tau_g\sim - (\gamma-\gamma_c) \text{ln} (\gamma-\gamma_c).
\label{10h}
\end{equation}

For $d>4$, as $\mathcal{I}'(\gamma)$ is convergent at the critical point and it immediately  follows that
\begin{equation}
\mathcal{I}(\gamma)=\mathcal{I}(\gamma_c)+(\gamma-\gamma_c)\mathcal{I}'(\gamma_c)+\cdots,
\end{equation}
which, in turn, when plugged in (\ref{9}) and expanded around the critical point, furnishes 
\begin{equation}
\tau_g\sim (\gamma-\gamma_c).
\label{10i}
\end{equation}

The above results can be summarized as 
\begin{eqnarray}
(\gamma-\gamma_c)\sim \left\{
\begin{array}
[c]{ccc}
\tau_g^{\frac{2}{d-2}} & \text{for} & 2<d<4\\
-\frac{\tau_g}{\text{ln}\tau_g}  &\text{for} & d=4\\
\tau_g &\text{for}& d>4
\end{array}
\right. ,
\label{15}
\end{eqnarray}
showing the behavior of $\gamma$ near the quantum critical point. This corresponds to a quantum phase transition without supersymmetry breaking.

\begin{figure}[h]
\includegraphics[width=7cm,height=6cm]{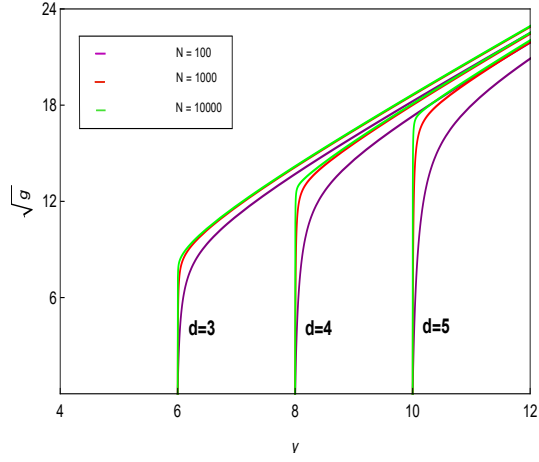}
\caption{Formation of the singularity according to the numerical analysis of Eq. (\ref{0.12}) with  $H_B=\mu=0$ as $N$ is increased. 
The zero mode (that leads the critical point for $\gamma>0$ and $U<0$) is treated separately before taking the 
thermodynamic limit.
For $U\equiv-1$, the critical points are $\sqrt{g_c} = 7.91$, $12.9$, and $17.29$, for dimensions 3, 4 and 5, respectively.}
\label{zero_temperature_N}
\end{figure}

A numerical analysis of Eq. (\ref{0.12}) can help reveal the relation between the parameters $\gamma$ and $\sqrt{g}$ as a function of $N$. The results are shown in Fig. \ref{zero_temperature_N} for some integer dimensions, where we observe the points of nonanalyticity arising as we increase $N$, i.e., as we go to the thermodynamic limit, signalling the quantum phase transition. Moreover, the numerical solution of Fig. \ref{zero_temperature_N} shows that $\sqrt{g}>\sqrt{g_c}$ for $\gamma>\gamma_c$, whereas $\sqrt{g}<\sqrt{g_c}$ for $\gamma=\gamma_c$. Notice that $\gamma$ never goes below the critical value $\gamma_c$.

\subsection{Magnetization, Fermionic Condensate, and Susceptibility}

In the quantum case, the thermodynamic quantities can be computed from the free energy (\ref{free_energy}) in the limit of zero temperature, with $\mu=0$, 
\begin{eqnarray}
f&=&-\frac{H_B^2}{2\left[U(0)+\gamma\right]^2}+
\frac{\sqrt{g}}{2N}\sum_{\bf q} 
\left\{
\left[\left(U({\bf q})+\gamma\right)^2\right]^{\frac{1}{2}}
-\left[\left(U({\bf q})+\gamma+H_F\right)^2\right]^{\frac{1}{2}}
\right\},
\label{17}
\end{eqnarray}
where $U(0)=-2|U|d=-\gamma_c$. We first compute the magnetization,
\begin{eqnarray}
m_{B} \equiv\left<\frac{1}{N} \sum_{\mathbf{r}} S_{\mathbf{r}}\right>
&=&- \frac{\partial f}{\partial H_{B}}=\frac{H_B}
{(\gamma-\gamma_c)^2},
\label{18}
\end{eqnarray}
which vanishes for $H_B=0$ and $\sqrt{g}>\sqrt{g_c}$, since $\gamma$ is always different from $\gamma_c$. However, when $H_B=0$ and $\sqrt{g}<\sqrt{g_c}$ there is an indeterminacy in the magnetization because, for such region, $\gamma =\gamma_c$.  In this case, we can use the spherical constraint in the presence of $H_B$ to settle this indeterminacy. Notice first that according to Eq. (\ref{9}), $g_c$ is given by 
\begin{eqnarray}
\frac{1}{\sqrt{g_c}}&=&\frac{1}{4|U|}\int_0^{\infty}dt\exp\left(-\frac{\gamma_c}{2|U|}t\right) \left[\text{I}_0(t)\right]^d\nonumber\\
&=&\frac{1}{4|U|}\int_0^{\infty}dt e^{-d t} \left[\text{I}_0(t)\right]^d.
\label{18a}
\end{eqnarray}
Now, considering the constraint equation for values $\sqrt{g}<\sqrt{g_c}$ and including the dependence on the external field $H_B$, we obtain
\begin{eqnarray}
1&=& \frac{H_B^2}{\left(\gamma-\gamma_c\right)^4}
+\frac{\sqrt{g}}{4|U|}\int_0^{\infty}dt\exp\left(-\frac{\gamma_c}{2|U|}t\right) \left[\text{I}_0(t)\right]^d\nonumber\\
&=&\frac{H_B^2}{\left(\gamma-\gamma_c\right)^4}+ \frac{\sqrt{g}}{\sqrt{g_c}}.
\label{19}
\end{eqnarray}
By using (\ref{18}) in this relation it follows immediately that
\begin{eqnarray}
m=\pm\left(\frac{\sqrt{g_c}-\sqrt{g}}{\sqrt{g_c}}\right)^{\frac{1}{2}},
\label{20}
\end{eqnarray}
giving the quantum critical exponent $\beta_g=1/2$ for $d>2$. As in the non-supersymmetric counterpart, the magnetization does not depend on the dimension \cite{Vojta1}. 

Analogously to the bosonic magnetization, the fermionic condensate can be computed as
\begin{eqnarray}
C_{F} \equiv \left<\frac{1}{N} \sum_{\mathbf{r}} \bar{\psi_{\mathbf{r}}} 
\psi_{\mathbf{r}}\right>
&=&- \frac{\partial f}{\partial H_{F}}\nonumber\\
&=&\frac{\sqrt{g}}{2}\frac{1}{N}\text{sign}\sum_{\bf q}
\left(\gamma+U({\bf q})+H_F\right)\nonumber\\
&=&\frac{\sqrt{g}}{2}\text{sign}(\gamma+H_F),
\label{21}
\end{eqnarray}
where we have used that $\sum_{\bf q}U({\bf q})=U(|{\bf h}|=0)=0$, since there is no self-interaction. Therefore, for $H_F=0$, the fermionic condensate is a function of $g$ both below and above the critical point.

Finally, we obtain the bosonic susceptibility from Eq. (\ref{18}),
\begin{eqnarray}
\chi_{B}=\frac{\partial m_{B}}{\partial H_{B}}
=\left(\gamma-\gamma_{c}\right)^{-2},
\label{22}
\end{eqnarray}
which diverges for $\sqrt{g}<\sqrt{g_c}$ since $\gamma=\gamma_c$. This is a characteristic of spherical models \cite{Joyce}. For $\sqrt{g}>\sqrt{g_c}$ the quantity $(\gamma-\gamma_c)$ depends on the dimension according to  Eq. (\ref{15}). In particular, for $2<d<4$, we get
\begin{eqnarray}
\chi_{B}=\tau_{g}^{-\frac{4}{d-2}},
\label{24}
\end{eqnarray}
so that we find the new critical exponent $\gamma_g=\frac{4}{d-2}$, showing that the supersymmetric quantum spherical model indeed exhibits a non-trivial behavior for the case of short-range interactions. It is instructive to compare this critical exponent with the non-supersymmetric counterpart, given by $\gamma_g=\frac{2}{d-2}$ \cite{Vojta1}.
 
Above the upper critical dimension, $d>4$, Eq. (\ref{15}) implies
\begin{eqnarray}
\chi \sim \tau_{g}^{-2},
\label{23}
\end{eqnarray}
recovering the mean-field critical exponent $\gamma_g=2$ \cite{Lucas}. 


\section{Thermal Critical Behavior}\label{S4}

In the previous case we have seen that supersymmetry is not spontaneously broken at zero temperature as the saddle point equations enforce $\mu=0$. At finite temperature, however, supersymmetry is always broken. This is a consequence of the different way in which bosons and fermions behave in the presence of thermal fluctuations \cite{Girardello,Daniel}. Therefore, it is expected in this case that the saddle point equations admit solutions with $\mu\neq 0$.  

At finite temperature the thermal fluctuations in general dominate over quantum fluctuations ($\beta w_{\bf q}^{B/F}<<1$), so that the critical behavior is governed essentially by the former. In this situation, we can expand the hyperbolic functions in Eqs. (\ref{0.12}) and (\ref{0.13}) for small arguments ($\coth x \sim \frac{1}{x}+\frac{x}{3}$ and $\tanh x\sim x$) to study the critical behavior. In the absence of external fields, we then find 
\begin{eqnarray}
1\approx\frac{1}{2\beta N}\sum_{\bf q}\frac{1}{ \left[\mu+\frac{1}{2}\left(\gamma+2U\sum_i\cos(q_i)\right)^2\right]},
\label{25}
\end{eqnarray}
and
\begin{eqnarray}
0&\approx& \frac{1}{2\beta N}\sum_{\bf q}\frac{1}{\left[\mu+\frac{1}{2}\left(\gamma+2U\sum_i\cos(q_i)\right)^2\right]}
\left[2U\sum_i\cos(q_i)+\gamma\right]\nonumber\\
&-&\frac{\beta g}{6N}\sum_{\bf q}\left[2U\sum_i\cos(q_i)+\gamma\right].
\label{26}
\end{eqnarray}
Eq. (\ref{25}) shows that the model can exhibit a critical behavior for the whole region $\mu\leq0$. Remembering that $\gamma>0$ and $U<0$, the critical point occurs at the minimum of $U({\bf q})$, which now reads 
\begin{equation}
\gamma_{c}= 2|U|d +\sqrt{2|\mu|}.
\label{27}
\end{equation}

The analysis of the critical behavior follows similarly as in the case of the zero temperature and  can be obtained from  Eq. (\ref{25}) in the thermodynamic limit,
\begin{eqnarray}
\beta&=&\int \frac{d^dq}{(2\pi)^d} \frac{1}{\left[-2|\mu|+\left[\gamma-2|U|\sum_i\cos(q_i)\right]^2\right]}\nonumber\\
&=& \frac{1}{2\sqrt{2|\mu|}}\int \frac{d^dq}{(2\pi)^d}\left[\frac{1}{|\gamma-2|U|\sum_i\cos(q_i)|-\sqrt{2|\mu|}}-\frac{1}{|\gamma-2|U|\sum_i\cos(q_i)|+\sqrt{2|\mu|}}\right]
\nonumber\\&=&\frac{1}{2|U|\sqrt{2|\mu|}}
\int_0^{\infty} dt \exp\left(-\frac{\gamma t}{2|U|}\right)\sinh\!\left(\frac{\sqrt{2|\mu|}}{2|U|}t\right)
\left[\text{I}_0(t)\right]^d,
\label{28}
\end{eqnarray}
where in the last line we have employed the representation in (\ref{8}). Analysing the asymptotic behaviors, it is straightforward to show that the integral converges when $t \rightarrow 0$ for any dimension regardless the value of $\mu$. On the other hand, for large values of $t$ the convergence depends on the dimension as well as on the parameter $\mu$ and we shall investigate the cases $\mu\neq0$ and $\mu=0$ separately.

\subsection{Solutions with $\mu\neq0$}

\begin{figure}[h]
\includegraphics[width=7cm,height=6cm]{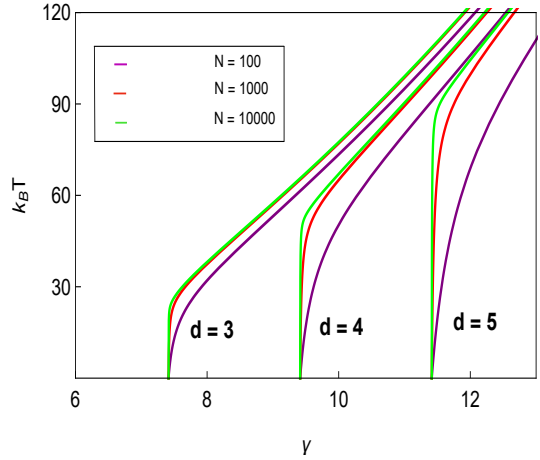}
\caption{Eq. (\ref{25}) leads to a nonanalyticity point as $N$ is increased. The plot was made using  $\mu=U=-1$ for $d=3,4,$ and $5$, defining the respective sets of critical points: 
($\gamma_c=7.41$; $k_BT_c=22.02$), ($\gamma_c=9.41$; $k_BT_c=51.69$), and ($\gamma_c=11.41$; $k_BT_c=88.71$).}
\label{temp_mu}
\end{figure}

For $\mu\neq 0$ the asymptotic behavior of the integral
\begin{eqnarray}
\mathcal{I}_d(\gamma,\mu)\equiv\int_0^{\infty} dt \exp\!\left(-\frac{\gamma }{2|U|}t\right)\sinh\!\left(\frac{\sqrt{2|\mu|}}{2|U|}t\right)
\left[\text{I}_0(t)\right]^d,
\label{29}
\end{eqnarray}
shows that divergences can occur for large values of $t$ depending on the dimensionality. In this case, the integral $\mathcal{I}_d(\gamma,\mu)$ is convergent for $d>2$ when $\gamma>\gamma_c$, exhibiting a critical point at $\gamma_c=2|U|d +\sqrt{2|\mu|}$. As in the previous section, we obtain the dependence of $(\gamma-\gamma_c)$ in the large-$t$ region from the derivative of Eq. (\ref{29}) with respect to $\gamma$,
\begin{eqnarray}
\mathcal{I}'_d(\gamma,\mu)\sim - \int_0^{\infty} dt ~t^{-\frac{(d-2)}{2}}\exp[-t(\gamma-\gamma_c)].
\label{30}
\end{eqnarray}
Comparing with the Eq. (\ref{10b}) we verify that for finite temperature and $\mu\neq0$ the model exhibits the same convergence properties as in the quantum case, showing a thermal phase transition with supersymmetry broken. According to Eq. (\ref{28}), the integral $\mathcal{I}_d(\gamma,\mu)$ is proportional to $1/T$, so that
\begin{eqnarray}
(\gamma-\gamma_c)\sim \left\{
\begin{array}
[c]{ccc}
\tau_{\beta}^{\frac{2}{d-2}} & \text{for} & 2<d<4\\
-\frac{\tau_{\beta}}{\text{ln}\tau_{\beta}}  &\text{for} & d=4\\
\tau_{\beta} &\text{for}& d>4
\end{array}
\right. ,
\label{31}
\end{eqnarray}
with $\tau_{\beta}\equiv (T-T_c)/T_c$. The points of non-analyticity at $\gamma=\gamma_{c}$ and how they depend  on the dimensions of the system 
are illustrated in Fig. \ref{temp_mu}.

\subsubsection{Magnetization, Fermionic Condensate, and Susceptibility}

Let us start with the bosonic magnetization, which is obtained from the free energy, Eq. (\ref{free_energy}), 
\begin{eqnarray}
m=-\frac{\partial f}{\partial H_B}=\frac{H_B}{2\left[-|\mu|+\frac{1}{2}\left(\gamma-2|U|d\right)^2\right]}.
\label{33}
\end{eqnarray}
According to the numerical solution shown in Fig. \ref{temp_mu}, when $T>T_c$ the parameter $\gamma$ is always greater than $\gamma_c$, such that the quantity 
$\left[-|\mu|+\frac{1}{2}\left(\gamma-2|U|d\right)^2\right]$ is different from zero. Thus, for $H_B=0$ the magnetization vanishes. For $T<T_c$, we have $\gamma=\gamma_c$ and the magnetization leads to an indeterminacy when $H_B=0$. As in the case of zero temperature, we can settle this by using the spherical constraint,  Eq. (\ref{25}), in the presence of $H_B$. With this we find
\begin{eqnarray}
m_B=\pm\left(\frac{T_c-T}{T_c}\right)^{\frac{1}{2}},
\label{34}
\end{eqnarray}
characterizing a thermal critical exponent $\beta_{T}=1/2$ for all dimensions $d>2$. 

\begin{figure}[h]
\includegraphics[width=7cm,height=6cm]{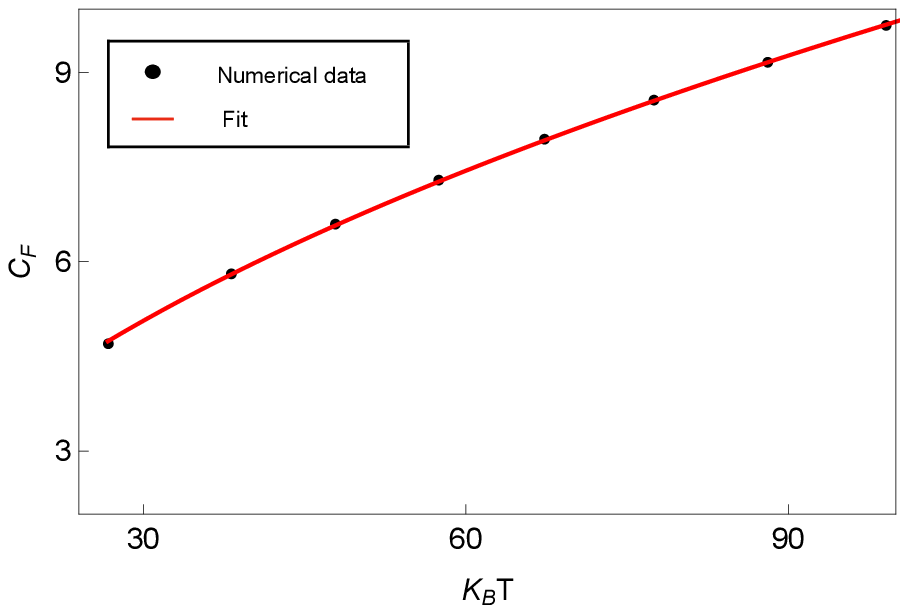}
\includegraphics[width=7cm,height=6cm]{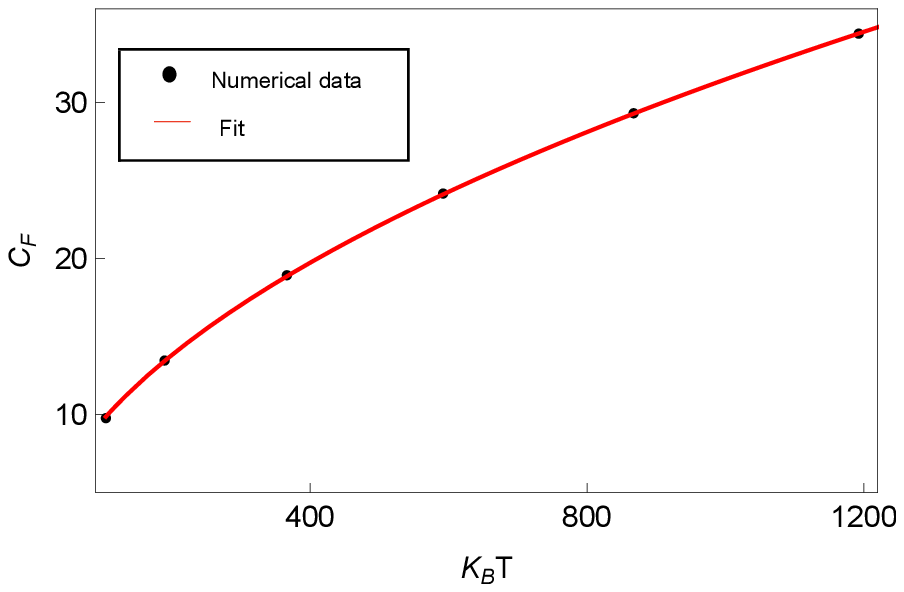}
\caption{The plots show a numerical analysis of the fermionic condensate, with $\mu\neq 0$, in the region $T>T_c$ 
for $d = 3$ and $d = 5$ in the left and right panels, respectively. The best fit of the numerical data is $C_F = a+b\sqrt{T}$, where 
$a$ and $b$ are constants.}
\label{fcmu}
\end{figure}

As the bosonic magnetization, from the free energy Eq. (\ref{free_energy}) we obtain the expression for the fermionic condensate,
\begin{eqnarray}
C_F=-\frac{\partial f}{\partial H_F}
=\frac{1}{\beta}\int \frac{d^dq}{(2\pi)^d}\frac{\gamma-2|U|\sum_i \cos(q_i)}{\left[-2|\mu|+\left(\gamma-2|U|\sum_i\cos(q_i)\right)^2\right]}.
\label{35}
\end{eqnarray}
In this expression, we have used the second constraint equation (\ref{26}) to express $g$ in terms of $\gamma$, $\mu$ and $\beta$. From this form, we see that for $T<T_c$, as $\gamma$ is fixed at $\gamma=\gamma_c$, the condensate behaves as 
\begin{eqnarray}
C_F\propto T,
\label{37}
\end{eqnarray}
independent of dimension. When $T>T_c$, $\gamma$ changes with the temperature but the expression (\ref{25}) does not furnish an explicit expression of $\gamma=\gamma(\beta)$. In this case, we proceed with a numerical analysis of the fermionic condensate in the region $\gamma>\gamma_c$. To this, we use the identity (\ref{8}) to rewrite (\ref{35})  in a more convenient way,
\begin{eqnarray}
C_F&=&\frac{1}{\beta}\frac{1}{2|U|\sqrt{2|\mu|}}\int_0^{\infty} dt\text{e}^{-\left(t\frac{\gamma}{2|U|}\right)}
\sinh\left(t\frac{\sqrt{2|\mu|}}{2|U|}\right)
\left[\gamma \left[\text{I}_0(t)\right]^d
-2|U| d \left[\text{I}_0(t)\right]^{d-1}
\text{I}_1(t)
\right].
\label{36}
\end{eqnarray}
The results are shown in Fig. \ref{fcmu}, where we see that above the critical temperature the condensate also does not depend on the dimension,
\begin{eqnarray}
C_F\propto T^{\frac{1}{2}}.
\label{38}
\end{eqnarray}
For very high temperatures we can see this behavior emerging in (\ref{35}). Indeed, in this limit the relation (\ref{25}) implies $\gamma\sim T^{\frac12}$ (neglecting $\mu$ and $U$ compared to $\gamma$), that when plugged into (\ref{35}) leads to the above result. Precisely these same dependencies with the temperature are obtained in the mean-field version of the model \cite{Lucas}.

Next we turn to the bosonic susceptibility from the Eq. (\ref{33}), 
\begin{eqnarray}
\chi_B=\frac{\partial m_B}{\partial H_B}=
\frac{1}{2}\left[
-|\mu|+\frac{1}{2}\left(\gamma-2d|U|\right)^2
\right]^{-1},
\label{39}
\end{eqnarray}
which diverges for $T<T_c$ since $\gamma=\gamma_c$. For $T>T_c$ we expand around the critical point, so that
\begin{eqnarray}
\chi_B &\sim & \frac{1}{2}\left[
\sqrt{2|\mu|}(\gamma-\gamma_c)+\frac{1}{2}(\gamma-\gamma_c)^2
\right]^{-1} .
\label{40}
\end{eqnarray}
As $\mu\neq 0$,  the  behavior of the susceptibility is governed by the dominant term $(\gamma-\gamma_c)$. 
Using Eq. (\ref{31}) to express   $(\gamma-\gamma_c)$  in terms of the temperature we finally obtain,
\begin{eqnarray}
\chi_{B} \sim \left\{
\begin{array}
[c]{ccc}
\tau_{\beta}^{-\left(\frac{2}{d-2}\right)} & \text{for} & 2<d<4\\
-\left(\frac{\tau_{\beta}}{\text{ln}\tau_{\beta}}\right)^{-1}  &\text{for} & d=4\\
\tau_{\beta}^{-1} &\text{for}& d>4
\end{array}
\right. .
\label{40.1}
\end{eqnarray}
For $2<d<4$ we obtain a new critical exponent, $\gamma_T=\left(\frac{2}{d-2}\right)$ and for $d>4$ we recover the mean-field exponent, $\gamma_T=1$  \cite{Lucas}.
  

\subsection{$\mu=0$}

By taking $\mu=0$ in (\ref{28}) we obtain
\begin{eqnarray}
\beta=\frac{1}{4|U|^2}\int_0^{\infty} dt ~t\exp\! \left(\!-\frac{\gamma}{2|U|}t\right)
\left[\text{I}_0(t)\right]^d.
\label{41}
\end{eqnarray} 
We see that the additional factor of $t$ in the integrand compared to the previous case will change the convergence properties and hence the critical behavior. Indeed, the integral we shall investigate now is
\begin{eqnarray}
\mathcal{I}_{d}^{0} \left(\gamma\right) \equiv
\int_0^{\infty} dt ~t\exp\! \left(\!-\frac{\gamma}{2|U|}t\right)
\left[\text{I}_0(t)\right]^d.
\label{1.41}
\end{eqnarray}
This integral converges at the critical point $\gamma=\gamma_{c}$ for $d>4$, which determine the lower critical dimension $d_{l}^{0}=4$. To determine the upper critical dimension we consider the derivative with respect to $\gamma$,
\begin{eqnarray}
\mathcal{I}_d^{'0}\left(\gamma\right) \sim -\int_{0}^{\infty} dt ~ t^{-\left(\frac{d-4}{2}\right)} \exp\left(-t\left(\gamma-\gamma_{c}\right)\right).
\label{2.41}
\end{eqnarray}
This expression converges at the critical point for $d>6$, giving the upper critical dimension $d_{u}^{0}=6$. By proceeding similarly as in Sec. \ref{S3}, we obtain the relation between ($\gamma-\gamma_{c}$) and $\tau_{\beta}$ according to the dimensionality,
\begin{eqnarray}
(\gamma-\gamma_c)\sim \left\{
\begin{array}
[c]{ccc}
\tau_{\beta}^{\frac{2}{d-4}} & \text{for} & 4<d<6\\
-\frac{\tau_{\beta}}{\text{ln}\tau_{\beta}}  &\text{for} & d=6\\
\tau_{\beta} &\text{for}& d>6
\end{array}
\right. .
\label{43}
\end{eqnarray}
Thus, the model exhibits a nontrivial critical behavior for $4<d<6$. 


\subsubsection{Magnetization, Fermionic Condensate, and Susceptibility}

\begin{figure}[h]
\includegraphics[width=7cm,height=6cm]{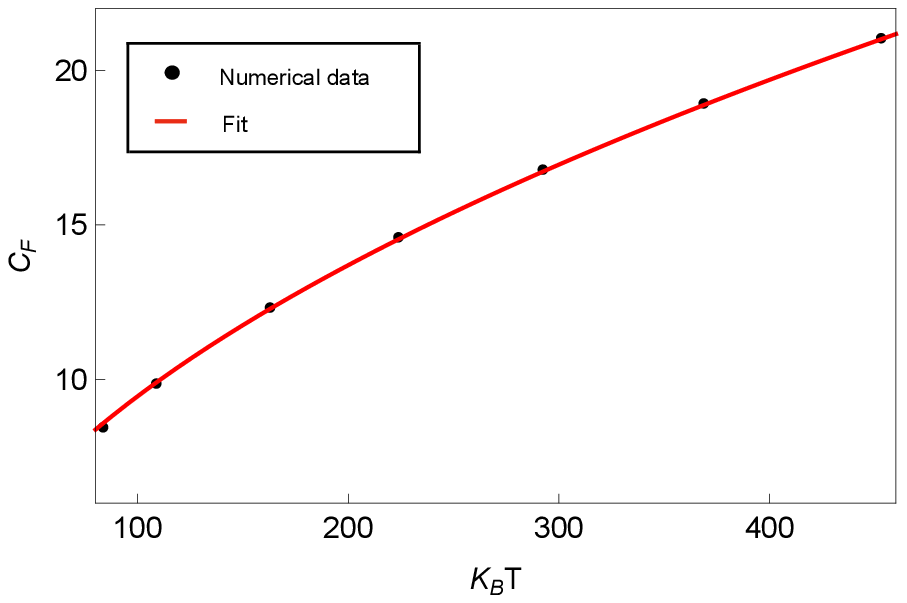}
\includegraphics[width=7cm,height=6cm]{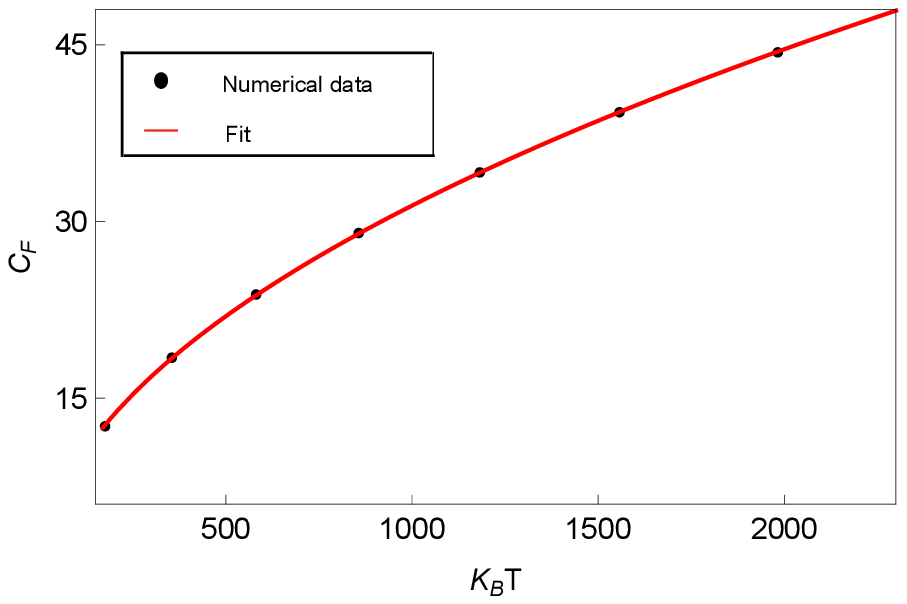}
\caption{Fermionic condensate for $\mu=0$ and $T>T_c$. The plots, for $d=5$ (left panel) and $d=7$ (right panel), show the same behaviors of the case
with $\mu\neq 0$. }
\label{cfmu0}
\end{figure}

Following the analysis of the previous sections, the critical exponents can be promptly computed. To avoid unnecessary repetition, we just quote the results. The critical exponent of the magnetization is $\beta_T=1/2$ for all dimensions $d>4$. The fermionic condensate also shows the same behavior as in the case $\mu\neq 0$, that is
\begin{eqnarray}
C_F\propto \left\{
\begin{array}
[c]{cc}
T & \text{for}~ T<T_c\\
T^{\frac{1}{2}}  & \text{for}~T>Tc  \\
\end{array}
\right.,
\label{fc0}
\end{eqnarray}
where the high-temperature dependence is determined numerically, as shown in Fig. {\ref{cfmu0}. The results are also independent of the dimension.

Finally, near the critical point the susceptibility behaves as 
\begin{eqnarray}
\chi_{B} \sim \left\{
\begin{array}
[c]{ccc}
\tau_{\beta}^{-\left(\frac{4}{d-4}\right)} & \text{for} & 4<d<6\\
\left(\frac{\tau_{\beta}}{\text{ln}\tau_{\beta}}\right)^{-2}  &\text{for} & d=6\\
\tau_{\beta}^{-2} &\text{for}& d>6
\end{array}
\right. ,
\label{40.1a}
\end{eqnarray}
which gives the critical exponents $\gamma_T=\frac{4}{d-4}$ for $4<d<6$, and $\gamma_T=2$ for $d>6$. 

A summary of the results is exhibit in the Table \ref{Table1}.
\begin{table}[h!]
\begin{tabular}{|c c c c|c c c c|c|}
\hline 
\hspace{.4cm} $T$ \hspace{.2cm} & \hspace{.2cm} $\mu$ \hspace{.2cm} & \hspace{.2cm} $d_{l}$ \hspace{.2cm}
& \hspace{.2cm} $d_{u}$\hspace{.4cm} & $\gamma_{g/T}$ \hspace{.4cm}& $\nu$ \hspace{.4cm}
& \hspace{.4cm} $\eta$\hspace{.4cm} 
& \hspace{.4cm}$\beta$ \hspace{.4cm} & univ. class \\ 
\hline 
 0 & 0 & 2 & 4 &  $\begin{array} [c]{ccc}   \begin{cases} 4/\left(d-2\ \right) \\ 2 \end{cases}  \end{array}$ 
 &$\begin{array} [c]{ccc}   \begin{cases} 1/\left(d-2\ \right) \\ 1/2 \end{cases}  \end{array}$ 
 & -2& $\frac{1}{2}$ & SUSY quantum \\ 
\hline $>0$ & $\neq 0$ & 2 & 4 & $\begin{array} [c]{ccc} \begin{cases} 2/ \left( d-2 \right) \\ 1  \end{cases} \end{array}$ 
&$\begin{array} [c]{ccc} \begin{cases} 1/ \left( d-2 \right) \\ 1/2  \end{cases} \end{array}$ 
&0&  $\frac{1}{2}$ & SUSY thermal I \\\hline 
$>0$ & 0 & 4 & 6 & $\begin{array} [c]{ccc} \begin{cases} 4/ \left( d-4 \right) \\ 2 \end{cases} \end{array}$ 
&$\begin{array} [c]{ccc} \begin{cases} 1/ \left( d-4 \right) \\ 1/2 \end{cases} \end{array}$ 
&-2&   $\frac{1}{2}$ & SUSY thermal II \\ 
\hline 
$0$ & - & 1 & 3 & $\begin{array} [c]{ccc} \begin{cases} 2/ \left( d-1 \right) \\ 1 \end{cases} \end{array}$ 
& $\begin{array} [c]{ccc} \begin{cases} 1/ \left( d-1 \right) \\ 1/2 \end{cases} \end{array}$ 
& 0& $\frac{1}{2}$ & quantum \\ 
\hline 
$>0$ & - & 2 & 4 & $\begin{array} [c]{ccc} \begin{cases} 2/ \left( d-2 \right) \\ 1 \end{cases} \end{array}$ 
&$\begin{array} [c]{ccc} \begin{cases} 1/ \left( d-2 \right) \\ 1 /2\end{cases} \end{array}$ 
& 0 & $\frac{1}{2}$ & classical  \\ 
\hline 
\end{tabular} 
\caption{Summary of the critical behavior. To facilitate comparison, we have included in the two bottom rows of the table the corresponding results for the non-supersymmetric quantum and classical versions of the spherical model. In addition, we have included the results for the critical exponents $\eta$ and $\nu$. The exponent $\eta$ follows directly from the behavior of the correlation function $\langle S_{-{\bf q}}S_{\bf q}\rangle$ for small momenta (large distances) and taken at the critical point, i.e., $\langle S_{-{\bf q}}S_{\bf q}\rangle\sim (w_{\bf q}^B)^{-2}\sim \left(-\sqrt{2|\mu|}U|{\bf q}|^2+ \frac12 U^2 |{\bf q}|^4+\cdots \right)^{-1} \sim|{\bf q}|^{-2+\eta}$, which shows clearly the difference of the values of $\eta$ in the cases of $\mu=0$ and $\mu\neq 0$. The exponent $\nu$ can be computed in all the cases from $\xi\sim(\gamma-\gamma_c)^{-\frac{1}{2}}$, and then using relations (\ref{15}), (\ref{31}) and (\ref{43}) to express $\xi$ in the form $\xi\sim \tau^{-\nu}$ for each one of the cases. We can check immediately that all the exponents satisfy the standard scaling relations, recalling that in the case of zero temperature we must replace $d\rightarrow d+z$.}
\label{Table1}
\end{table}


\section{Correlation Function}\label{S5}

The on-shell form of the Lagrangian (\ref{2.19a}) reveals an interesting feature of interactions between the bosonic variables $S_{\bf r}$ of different sites. The on-shell formulation is obtained by integrating out the auxiliary field $F_{\bf r}$. As it appears only quadratically in the action, this process is equivalent to use its equation of motion, which is purely algebraic
\begin{equation}
\frac{\partial L}{\partial F_{\bf r}}=0~~~\Rightarrow~~~ F_{\textbf{r}} 
= -\gamma S_{\textbf{r}}-\sum_{\textbf{r}'}U_{\textbf{r},\textbf{r}'}S_{\textbf{r}'},
\label{3.2}
\end{equation}
i.e., there is no time derivative of $F_{\bf r}$ and hence it is not a dynamical physical degree of freedom. Plugging this back in (\ref{2.19a}), we  obtain the on-shell Lagrangian
\begin{eqnarray}
L & = & \frac{1}{2}\sum_{\mathbf{r}}\dot{S}_{\mathbf{r}}^{2}
+i\sum_{\mathbf{\mathbf{r}}}\bar{\psi}_{\mathbf{r}}\dot{\psi}_{\mathbf{r}}
-\frac{1}{2}\sum_{\textbf{r},\textbf{r}'}J_{\textbf{r},\textbf{r}'}S_{\textbf{r}}S_{\textbf{r}'}
-\sum_{\mathbf{r},\mathbf{r}'}U_{\mathbf{r},\mathbf{r}'}\bar{\psi}_{\mathbf{r}}\psi_{\mathbf{r}'}
- \sum_{\mathbf{r}}S_{\mathbf{r}}\left(\bar{\psi}_{\mathbf{r}}\xi+\bar{\xi}\psi_{\mathbf{r}}\right)\nonumber \\
 & -&\mu\sum_{\mathbf{r}}\left(S_{\mathbf{r}}^{2}-N\right)-\frac{1}{2}\gamma^{2}N
 -\gamma\left(\sum_{\textbf{r},\textbf{r}'}U_{\textbf{r},\textbf{r}'}S_{\textbf{r}}S_{\textbf{r}'}
 +\sum_{\mathbf{r}}\bar{\psi}_{\mathbf{r}}\psi_{\mathbf{r}}\right),
 \label{3.4}
\end{eqnarray}
where $J_{\textbf{r},\textbf{r}'}\equiv\sum_{\textbf{r}''}U_{\textbf{r},\textbf{r}''}U_{\textbf{r}'',\textbf{r}'}$.

Let us focus on the terms involving interactions of $S_{\bf r}$ in different sites, 
\begin{eqnarray}
L_{SS}\equiv-\frac{1}{2}\sum_{\textbf{r},\textbf{r}'}J_{\textbf{r},\textbf{r}'}S_{\textbf{r}}S_{\textbf{r}'}-\gamma\sum_{\textbf{r},\textbf{r}'}U_{\textbf{r},\textbf{r}'}S_{\textbf{r}}S_{\textbf{r}'}.
\label{ss}
\end{eqnarray}
To make clear the role of these terms, we consider explicitly the interaction (\ref{0.5}) for a one-dimensional lattice. In this case, we have $U_{r,r'}=U(\delta_{r,r'+1}+\delta_{r,r'-1})$ and $J_{r,r'}=U^2(\delta_{r,r'+2}+\delta_{r,r'}+\delta_{r,r'-2})$, so that (\ref{ss}) becomes 
\begin{equation}
L_{SS}=-\frac{U^2}{2}\sum_{r}(S_{r}S_{r+2}+S_{r}S_{r-2})-\gamma U\sum_{r}(S_{r}S_{r+1}+S_{r}S_{r-1}),
\label{ssa}
\end{equation}
with the Lagrange multiplier $\gamma$ playing the role of an interaction energy between first neighbors. Thus, we see that even if $U_{\textbf{r},\textbf{r}'}$ is only a nearest neighbor interaction, once we integrate out the auxiliary field $F_{\bf{r}}$, the resulting interactions $J_{\textbf{r},\textbf{r}'}$ will be effectively of second nearest-neighbors.
As $\gamma$ and $U$ have opposite signs, the interaction energies $U^2/2>0$ and $\gamma U<0$ favour different orderings between first and second neighbors (ferro and anti-ferro, respectively). In general, models with competing interactions give rise to rich phase diagrams with modulated phases, as it has been observed in several lattice models \cite{Selke,Selke1,Nussinov1,Xu}, including the classical \cite{Pisani,Henkel2} and quantum spherical models \cite{Paula_Salinas,Henkel1}. The competing interactions also affect the correlation functions. When the interactions are independent of each other, depending on their relative magnitude, they usually lead to an oscillatory behavior in the correlation, besides the usual exponential decay. In particular, as it has been shown in \cite{Paula_Salinas} for the quantum spherical model with competing interactions, such oscillation manifests already in the one-dimensional correlation function.

In the present case, however, the interactions are not independent at all, since the saddle point solution for $\gamma$ implies that it must satisfy the spherical constraints, which in turn involve the other parameters, including $U$. Therefore it is not clear a priori whether the correlation functions will exhibit oscillatory behavior. We investigate this point further by computing  the correlation function in the one-dimensional supersymmetric model.

The correlation function can be computed in the usual way by introducing a site-dependent field, through the replacement,
\begin{eqnarray}
H_{B}\sum_{{\bf r}}S_{{\bf r}}~ \Rightarrow~ \sum_{{\bf r}} H_{{\bf r}}S_{{\bf r}}.
\end{eqnarray}
With this, the free energy becomes
\begin{eqnarray}
f&=&-
\frac{1}{4N}\sum_{{\bf q}} \left( 2g \right) \frac{
H_{{\bf q}}  H_{-{\bf q}}
}{\left(
w_{{\bf q}}^{B}
\right)^{2}}
-\mu
+\frac{1}{\beta N}\sum_{\bf q}\ln\left[
2\sinh\left(\frac{\beta}{2}w_{\bf q}^B\right)
\right]\nonumber\\
&-&\frac{1}{\beta N}
\sum_{\bf q}\ln\left[
2\cosh\left(
\frac{\beta}{2}w_{\bf q}^F\right)
\right].
\label{free_energy2} 
\end{eqnarray}
In the momentum space, the correlation function follows immediately, 
\begin{eqnarray}
\left<
S_{{\bf q}}S_{-{\bf q} } 
\right>
= -
\frac{1}{\beta} \frac{\partial^{2}f}{\partial H_{{\bf q}}H_{-{\bf q}}}
= \frac{g}{\beta N}\frac{1}{\left(
w_{{\bf q}}^{B}
\right)^{2}}.
\label{Correlation_1a}
\end{eqnarray}
Turning back to the position space, the one-dimensional correlation function reads,
\begin{eqnarray}
\left<
S_{r}S_{r+h}
 \right>
= 
  \frac{1}{2 \beta}\int_{-\pi}^{\pi} \frac{dq}{2\pi} \frac{e^{ iqh }}{\mu+\frac12\left(
\gamma- 2|U| \cos q
\right)^{2}
}.
\label{Correlation1b}
\end{eqnarray}
For simplicity, in the following analysis we shall consider the saddle point solution with $\mu=0$, since it is not a relevant parameter for the question we are investigating. Of course, the general conclusions are not affected when $\mu\neq 0$. We have to analyse the correlation function together the constraint equation, 
\begin{eqnarray}
1 = 
\frac{1}{\beta}
\int_{-\pi}^{\pi} \frac{dq}{2\pi} \frac{1}{\left(
\gamma- 2|U| \cos q
\right)^{2}
}.
\label{onedimensional}
\end{eqnarray}

To make transparent the main point of this analysis, we shall consider a slightly more general computation, by deforming the correlation function according to
\begin{eqnarray}
\left<
S_{r}S_{r+h} 
\right>
= \frac{1}{\beta}\int_{-\pi}^{\pi} \frac{dq}{2\pi} \frac{\exp{ \left(iqh \right)}}{\left(
\gamma- 2|U| \cos q
\right)^{2}
+\alpha \cos q
},
\label{Correlation_3}
\end{eqnarray}
with the corresponding modification in the constraint (\ref{onedimensional}). This is equivalent to adding to the Lagrangian an independent first neighbor interaction with energy $\alpha$. Of course, adding only this term in the model is incompatible with supersymmetry\footnote{It would be possible in principle to modify the Lagrangian as well as the supersymmetry transformations in order to accommodate such a term without breaking supersymmetry. We will return to this point in the next section where we discuss a similar deformation without destroying supersymmetry in an equivalent field theoretical model. }. In the end we will take the limit $\alpha\rightarrow0$. 

With the change of variable $z=e^{iq}$, we trade the integration in $q$ by an integration along a closed path (unit circle) in the complex plane. The denominator of (\ref{Correlation_3}) can be written as a fourth-order polynomial in $z$, 
\begin{eqnarray}
\left<
S_{r}S_{r+h} 
\right>
= \frac{1}{2\pi i \beta |U|^{2}}
\oint\limits_{\mathcal{C}} dz \frac{z^{1+h}}{\left[
z^{4}
+ \left(\frac{\alpha}{2|U|^2}-
\frac{2\gamma}{|U|}
 \right)z^{3} 
 +\left(
 \frac{\gamma^{2}}{|U|^{2}} +2
 \right)z^{2}
 +\left(\frac{\alpha}{2|U|^2}
-\frac{2\gamma}{|U|}
 \right)z
 +1
\right]},
\label{Correlation2}
\end{eqnarray}
where the contour $\mathcal{C}$ is the unit circle travelled in the counterclockwise way. The roots of the polynomial are given by
\begin{eqnarray}
z_{1} = \frac{1}{8|U|^{2}}\left(
-\alpha +
4|U|\gamma - A - B
\right),~~~
z_{2} =  \frac{1}{8|U|^{2}} 
\left(
- \alpha +4|U|\gamma +A + B^{*}
\right),
\label{roots}
\end{eqnarray}
and the corresponding complex conjugates, $z_1^*$ and $z_2^*$, with
\begin{eqnarray}
A &\equiv & \sqrt{\alpha \left( 
\alpha - 8|U|\gamma
\right)}
,\nonumber \\
B &\equiv & \sqrt{-
8|U|\gamma \left[
A+2\left(\alpha - |U|\gamma \right)
\right]
-64|U|^{4}
+2\alpha \left(\alpha + A
\right)
}.
\end{eqnarray}
Out of the four roots, only $z_1$ and $z_1^*$ reside inside the unit circle if we consider values of $\alpha$ in the interval\footnote{We impose this upper bound on $\alpha$
just to simplify the analysis. What really matters here is that the value $\alpha=0$ belongs to the interval, which is the limit we intend to take as to recover the original model.} $0\leq\alpha<4|U|\gamma$. Computing the integration in the complex plane we obtain,
\begin{eqnarray}
\left<
S_{r}S_{r+h} 
\right>
=
\frac{1}{|U|^{2}\beta}
\left[
\frac{z_{1}^{(h+1)}}
{w^{}\left(\alpha, \gamma, U \right)}
+
\frac{z^{*(h+1)}_{1}}
{w^{*}\left(\alpha, \gamma, U \right)}
\right],
\label{Correlation3}
\end{eqnarray}
where we have written the denominator as
\begin{eqnarray}
w(\alpha, \gamma, U) \equiv (z_{1} - z_{2})
(z_{1}-z_{1}^{*})
(z_{1}-z_{2}^{*})
\equiv |w| e^{i \delta}.
\end{eqnarray}
The integral of the constraint (\ref{onedimensional}) (with the introduction of the $\alpha\cos q$ term) can be carried out similarly, resulting in
\begin{equation}
1=\frac{1}{|U|^2\beta}\left[
\frac{z_{1}}
{w^{}\left(\alpha, \gamma, U \right)}
+
\frac{z^{*}_{1}}
{w^{*}\left(\alpha, \gamma, U \right)}
\right].
\end{equation}
With this, we can write the correlation function (\ref{Correlation3}) in the compact form,
\begin{eqnarray}
\left<
S_{r}S_{r+h} 
\right>
=
\frac{\cos
\left[
\delta -\left(h+1 \right)\theta
\right]
}{\cos \left(
\delta-\theta
\right)}
\exp
\left(
h \ln |z_{1}|
\right)
\label{Correlation_4},
\end{eqnarray}
where,
\begin{eqnarray}
z_{1}\equiv|z_{1}|e^{i\theta}.
\end{eqnarray}
We see that in the limit $\alpha\rightarrow 0$ the root $z_1$ becomes real so that $\theta$ vanishes. In this case, the oscillatory factor cancels out the correlation function (\ref{Correlation_4}), remaining only the usual exponential decay, since $|z_1|<1$ (this root is inside the unit circle). Therefore, the model does not exhibit the oscillations characteristic of the competing interactions. They are only generated when we deform the theory with the $\alpha\cos q$ term, which in turn corresponds to introducing an independent first neighbor interaction in the model. 



\section{Equivalence with the Nonlinear Sigma Model}\label{S6}

In this last part of our work, we discuss the equivalence with a field theory model. A useful guide to enlighten this connection is to recall the classical-quantum mapping based on renormalization-group arguments, connecting classical (thermal) critical phenomena in $D$ spatial dimensions to quantum critical phenomena (zero temperature) in $d=D-z$ spatial dimensions. In our case, we notice a shift of both lower and upper critical dimensions by a factor of 2; i.e., , $(2,4)\rightarrow(4,6)$, for the cases of zero temperature and finite temperature with $\mu=0$, showing that the dynamical critical exponent is $z=2$. It can also be obtained directly from the divergence of the correlation time at the critical point, $\tau_c\sim (w_{\bf q}^B)^{-1}\sim \left(-\sqrt{2|\mu|}U|{\bf q}|^2+ \frac12 U^2 |{\bf q}|^4+\cdots \right)^{-\frac12}\sim |{\bf q}|^{-z} $, which for $\mu=0$, gives $z=2$.
This implies that the correlation length and the correlation time scale anisotropically in the model, weighted by the exponent $z=2$, and hence any field theory connection should be with a nonrelativistic one. This feature can also be noticed even more directly by considering naively the continuum limit of the action in the superspace (\ref{0.1}) in the case where $U_{{\bf r},{\bf r}'}$  corresponds to first neighbor interactions. While the time derivatives appear only inside the supercovariant derivatives, the spatial derivatives emerging in the continuum limit appear explicitly in the action, 
\begin{equation}
\int dt d\theta d\bar{\theta} \sum_{{\bf r},{\bf r}'} U_{{\bf r},{\bf r}'}\Phi_{\bf r}\Phi_{{\bf r}'}~~\underset{\text{continuum limit}}{\longrightarrow}~~ \int dt d\theta d\bar{\theta}\int d^dr\left( \Phi_{\bf r}^2-\frac{1}{2} ({\nabla}\Phi_{\bf r})^2\right),
\label{9.0a}
\end{equation} 
and are thus not on equal footing with the time derivatives. The result is a theory that when written in components has a different number of temporal and spatial derivatives, leading to an anisotropic scaling weighted by $z=2$. 

Matching of symmetries is crucial in identifying the equivalent theory. According to the previous discussion, we should then look for a theory which is supersymmetric but not Lorentz invariant. A nonlinear sigma model with those properties has been constructed in \cite{Gomes7}, and is a natural candidate to be equivalent to the supersymmetric quantum spherical model. In addition, the nonlinear sigma model has an $O(N)$ internal symmetry that is not present in the spherical model. That is the reason why the equivalence will be established strictly in the limit $N\rightarrow \infty$, where effectively there is no symmetry at all. This also happens in all the theories in the right hand side of (\ref{9.0}).

For convenience, before going through the equivalence, we briefly review the construction of the model by following the conventions of \cite{Gomes7}.  As the spinor size and consequently the superspace structure depend on the spacetime dimension, we shall consider here explicitly the of $2+1$ spacetime dimensions, which can also be used for the case of $1+1$ dimensions, where we have for both situations two-component spinors.  In this case, the superspace is constituted of bosonic coordinates $x^{0}$ and $x^{i}$, with $i=1,2$, and a pair of real Grassmann coordinates $\theta_{\alpha}$, with $\alpha=1,2$. 

The model is constructed out of a set of $N$ scalar superfields\footnote{Although we are using the same letter $N$ in both models, it has different meanings in each case. In the spherical model, the thermodynamic limit necessarily corresponds to $N\rightarrow\infty$, since there is represents the total number of sites of the lattice. On the other hand,  it is a free parameter in the nonlinear sigma model, i.e., the number of fields, that can be chosen at our convenience.}, 
\begin{equation}
\Phi_a=\varphi_a+\bar\theta\psi_a+\frac{1}{2}\bar{\theta}\theta F_a,~~~a=1,...,N,
\label{9.1}
\end{equation} 
having as components, real scalar fields, $\varphi_a$, Majorana spinor fields, $\psi_a$, and auxiliary bosonic fields $F_a$. The conjugated Grassmann variable is defined as  
$\bar \theta\equiv \theta^{T}\gamma^{0}$ and similarly for $\bar\psi$. The Dirac matrices in the $2\times 2$ representation are given in terms of the Pauli matrices as $\gamma^0=\sigma_2$, $\gamma^1=i\sigma_1$ and $\gamma^2=i\sigma_3$.

The superfields are required to satisfy the constraint
\begin{equation}
\Phi_a\Phi_a=\frac{N}{2\tilde{g}},
\label{9.2}
\end{equation}
with $\tilde{g}$ being the coupling constant. It is important to emphasize the fundamental difference compared with the spherical constraint (\ref{0.1a}). While the above constraint is local, since it involves only fields at the same point of the spacetime, the spherical constraint (\ref{0.1a}) involves spin variables of all sites of the lattice, even those which are far apart from each other.

In terms of component fields, the constraint reads,
\begin{equation}
\varphi_a\varphi_a=\frac{N}{2\tilde{g}},~~~\psi_a\varphi_a=0,~~~\text{and}~~~\varphi_a F_a=\frac12 \bar\psi_a\psi_a.
\label{9.3}
\end{equation}
These constraints can be imposed via a superfield Lagrange multiplier
\begin{equation}
\Sigma=\sigma+\bar{\theta}\xi+\frac{1}{2}\bar{\theta}\theta\lambda,
\label{9.4}
\end{equation}
through the inclusion of the term $\Sigma(\Phi_a\Phi_a-N/2\tilde{g})$ in the action.

The action of the model, incorporating the anisotropic scaling characterized by $z=2$, is given by 
\begin{equation}
S=\frac12\int dt d^2x d^2\theta \left[\Phi_a \bar{D}{D}\Phi_a+ a_2\Phi_a \nabla^2\Phi_a   -
\Sigma (\Phi_a\Phi_a-\frac{N}{2\tilde{g}})\right],
\label{9.5}
\end{equation}
where $a_2$ is a dimensionless positive parameter and the supercovariant derivative is defined as 
\begin{equation}
D\equiv \frac{\partial}{\partial\theta}-i \bar\theta\gamma^{\mu} \tilde{\partial}_{\mu},~~~(\tilde{\partial}_{\mu}\equiv \partial_0,\,a_1\partial_i).
\label{9.6}
\end{equation}
Here $a_1$ is a dimensionfull parameter ($[a_1]=1$ in mass units) to give the correct dimension for the supercovariant derivative, since for the case $z=2$ it is more convenient to assign dimensions $[x^0]=2$ and $[x^i]=1$ in length units.

The supersymmetry transformations can be obtained from the supercharge,  
\begin{equation}
Q\equiv \frac{\partial}{\partial\bar\theta}+i \gamma^{\mu}\theta \tilde{\partial}_{\mu},
\label{9.7}
\end{equation}
defined in such way that it anticommutes with the supercovariant derivative, i.e., $\{D,Q\}=0$. This is required in order to ensure that $D\Phi_a$ transforms as the superfield itself, so that any term in the superspace action involving supercovariant derivatives is manifestly supersymmetric. The supercharge generates translations in superspace,
\begin{equation}
\delta x^0\equiv \bar\epsilon Q x^0=i\bar\epsilon \gamma^0 \theta,~~~
\delta x^i\equiv \bar\epsilon Q x^i=ia_1 \bar\epsilon \gamma^i \theta~~~\text{and}~~~\delta \theta\equiv \bar\epsilon Q \theta=\epsilon,
\label{9.8}
\end{equation}
where $\epsilon_{\alpha}$, $\alpha=1,2$, is a Grassmannian parameter of the transformation, under which the superfield transforms as 
\begin{equation}
\delta\Phi_a\equiv \bar\epsilon Q\Phi_a.
\label{9.9}
\end{equation}
By using the properties $\bar{\theta}_{\alpha}\theta_{\beta}=i \theta _{2}\theta_{1}\delta_{\alpha\beta}=\frac{1}{2}\bar{\theta}\theta\delta_{\alpha\beta}$ and 
$\bar{\epsilon}\theta=\bar{\theta}\epsilon$, and comparing the corresponding power of $\theta$ in both sides, we find the supersymmetry transformations of the components
\begin{eqnarray}
\delta\varphi_a&=& \bar\epsilon\psi_a,\nonumber\\
\delta \psi_a&=& -i\gamma^{\mu}\epsilon\tilde{\partial}_{\mu}\varphi_a+F_a\epsilon,\nonumber\\
\delta F_a&=& -i\bar{\epsilon} \gamma^{\mu}\tilde{\partial}_{\mu}\psi_a. 
\label{9.10}
\end{eqnarray}


\subsection{Action in Components and the Large $N$ Expansion}

Now we are ready to discuss the equivalence between the models. To show this it is convenient to write the action (\ref{9.5}) in terms of components, 
\begin{eqnarray}
S&=&\int dt d^2x\left[-\frac12 \varphi\tilde{\partial}^2\varphi+\frac{i}{2}\bar\psi\gamma^{\mu}\tilde{\partial}_{\mu}\psi+
\frac12 F^2
-a_2 F\nabla^2\varphi+\frac{a_2}{2}\bar\psi\nabla^2\psi\right.\nonumber\\
&+&\left. \sigma(F\varphi-\frac12\bar\psi\psi)-\bar\xi\psi\varphi+
\frac{\lambda}{2}(\varphi^2-\frac{N}{2\tilde{g}})\right].
\label{9.11}
\end{eqnarray}
Next we use the equation of motion of the auxiliary field $F_a$,
\begin{equation}
F_a=a_2\nabla^2\varphi_a-\sigma\varphi_a,
\label{9.12}
\end{equation}
to eliminate it from the Lagrangian,
\begin{eqnarray}
S&=&\int dt d^2x\left[-\frac12 \varphi\tilde{\partial}^2\varphi-\frac{a_2^2}{2}\nabla^2\varphi\nabla^2\varphi+
\frac{i}{2}\bar\psi\gamma^{\mu}\tilde{\partial}_{\mu}\psi
+\frac{a_2}{2}\bar\psi\nabla^2\psi\right.\nonumber\\
&+&\left. a_2\sigma\varphi\nabla^2\varphi-\frac{1}{2}\sigma^2\varphi^2-\frac12\sigma\bar\psi\psi-\bar\xi\psi\varphi+
\frac{\lambda}{2}(\varphi^2-\frac{N}{2\tilde{g}})\right].
\label{9.13}
\end{eqnarray}
Spontaneous supersymmetry breaking is related to the possibility that the Lagrange multiplier fields acquire nonvanishing vacuum expectation value. To appreciate this point, we make the shifts $\sigma \rightarrow \sigma+m^2$ and $\lambda\rightarrow\lambda+\lambda_0$, where $m^2$ and $\lambda_0$ are the vacuum expectation value of the fields $\sigma$ and $\lambda$, i.e., $\langle \sigma\rangle\equiv m^2$ and $\langle\lambda\rangle\equiv \lambda_0$. Rotational invariance implies $\langle \xi \rangle=0$. These shifts provide masses for bosons and fermions,
\begin{equation}
\mathcal{L}_{mass}=-\frac{1}{2}(m^4-\lambda_0)\varphi^2-\frac{m^2}{2}\bar{\psi}\psi.
\label{9.14}
\end{equation}
Thus, whenever $\lambda_0\neq 0$, spontaneous supersymmetry breaking takes place since it induces an imbalance between the boson and fermion masses, independent of the value of $m^2$.

As the dependence on the fields $\varphi$ and $\psi,\bar{\psi}$ is at the most quadratic, they can be integrated out in the partition function. This produces an effective action that can be arranged in an expansion in powers of $1/N$. To this end, after the above shifts, we make an appropriate rescaling of the Lagrange multiplier fields, $\sigma\rightarrow \sigma/\sqrt{N}$, $\lambda\rightarrow\lambda/\sqrt{N}$ and $\xi\rightarrow \xi/\sqrt{N}$, which affect the interaction terms in the second line of (\ref{9.13}),
\begin{equation}
\mathcal{L}_{int}= \frac{1}{\sqrt{N}}\sigma\varphi(a_2\nabla^2-m^2)\varphi-\frac{1}{2N}\sigma^2\varphi^2-\frac{1}{2\sqrt{N}}\sigma\bar\psi\psi-\frac{1}{\sqrt{N}}\bar\xi\psi\varphi+
\frac{\lambda}{2\sqrt{N}}(\varphi^2-\frac{N}{2\tilde{g}}).
\label{9.15}
\end{equation}
The corresponding Feynman rules of the theory are depicted in Fig. \ref{lines} and Fig. \ref{vertices}. In addition to the factors of $N$ coming from the vertices, whenever there is a bosonic or fermionic loop this produces a factor of $N$ in the numerator, as the loop is produced by contracting $N$ fields. The fermionic propagator is 
\begin{eqnarray}
S_{ab}(p)&=& \frac{i\delta_{ab}}{\tilde{p}_{\mu}\gamma^{\mu}-(a_2 {\bf p}^{2}+m^{2})+i \epsilon}=
i\delta_{ab} \frac{\tilde{p}_{\mu}\gamma^{\mu}+(a_2 {\bf p}^{2}+m^{2})}{\tilde{p}^{2}-(a_2 {\bf p}^{2}+m^{2})^2+i \epsilon},
\label{9.16}
\end{eqnarray}
and the bosonic one reads
\begin{equation}
\Delta_{ab}(p)=  \frac{i\delta_{ab}}{\tilde{p}^{2}-(a_2{\bf p}^{2}+m^{2})^{2}+\lambda_0+i\epsilon}.
\label{9.17}
\end{equation}
\begin{figure}[h]
\includegraphics[scale=.7]{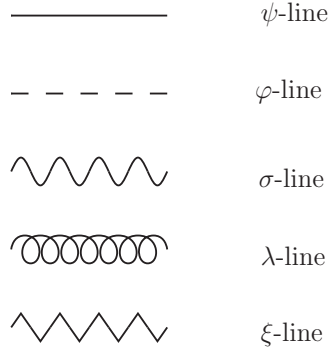}
\caption{Feynman rules - type of lines.}
\label{lines}
\end{figure}
\begin{figure}[h]
\includegraphics[scale=.57]{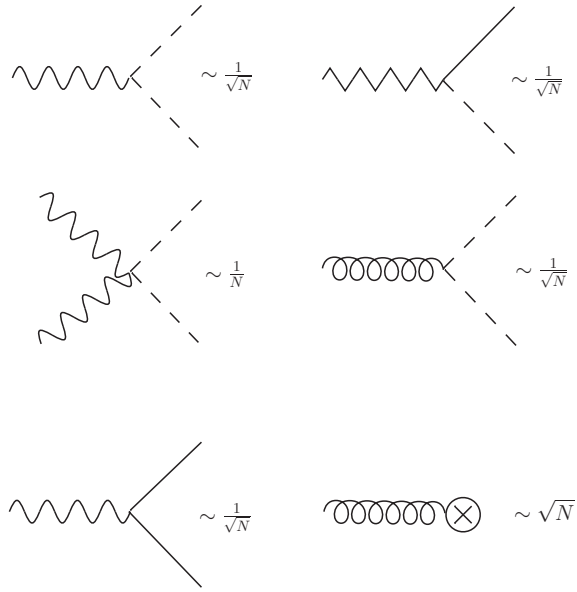}
\caption{Interaction vertices obtained from (\ref{9.15}).}
\label{vertices}
\end{figure}

The effective action obtained upon integration over the fields $\varphi$, $\psi$, and $\bar{\psi}$ has the structure of an $1/N$ expansion, 
\begin{equation}
S_{eff}[\sigma,\lambda,\xi]= N^{\frac12} S_1+ N^0 S_{2}+O(N^{-\frac12}),
\label{9.18}
\end{equation}
where $S_1$ represents the one-point functions of $\sigma$ and $\lambda$, 
\begin{equation}
S_1=\int dx \Gamma^{(1)}_{\sigma}(x=0)\sigma(x)+\int dx \Gamma^{(1)}_{\lambda}(x=0)\lambda(x),
\label{9.18a}
\end{equation}
with $dx\equiv dx^0d^dx$ and $\Gamma^{(1)}_{\sigma}(x=0)$ and $\Gamma^{(1)}_{\lambda}(x=0)$ given by the corresponding 1PI diagrams of Fig. \ref{oneloop}, which are of order $N^{\frac{1}{2}}$. Notice that we have extracted the factor of $N^{\frac12}$ to exhibit it explicitly in (\ref{9.18}). The Gaussian contribution $S_{2}$ involves the two-point functions, 
\begin{eqnarray}
S_2&=&-\frac{i}{2} \int dx dy \sigma(x)\Gamma^{(2)}_{\sigma\sigma}(x-y)\sigma(y)-\frac{i}{2} \int dx dy \lambda(x)\Gamma^{(2)}_{\lambda\lambda}(x-y)\lambda(y)\nonumber\\
&-&\frac{i}{2} \int dx dy \bar{\xi}(x)\Gamma^{(2)}_{\xi\xi}(x-y)\xi(y),
\label{9.18b}
\end{eqnarray}
with $\Gamma^{(2)}$'s given by the corresponding 1PI diagrams of Fig. \ref{twopoint}.
\begin{figure}[h]
\includegraphics[scale=.6]{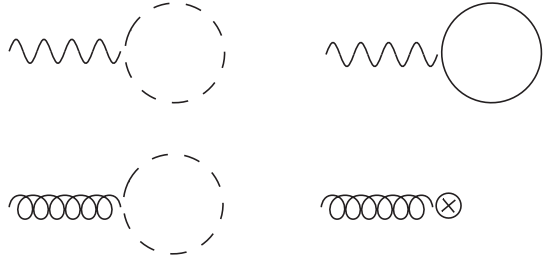}
\caption{One-loop contributions to the 1PI one-point functions. All these diagrams are of order $\sqrt{N}$.}
\label{oneloop}
\end{figure}

\begin{figure}[h]
\includegraphics[scale=.6]{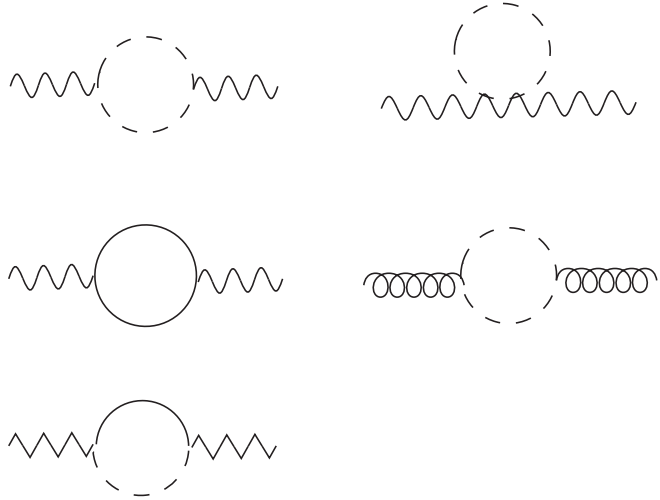}
\caption{One-loop contributions to the 1PI two-point functions. All these diagrams are of order $N^0$. Only these diagrams survive in the strict large $N$ limit.}
\label{twopoint}
\end{figure}

In order to have a well defined large $N$ expansion we shall impose that the contributions of $S_1$ vanish. Therefore, in the strict limit $N\rightarrow\infty$ the effective action is dominated by the saddle point given by the quadratic contribution $S_2$, precisely as in the case of the partition function of the supersymmetric quantum spherical model (\ref{0.7}). Now it remains to identify the parameters of the two theories. We can do this simply by analyzing the contributions in $S_1$. The sum of the two diagrams in the top of Fig. \ref{oneloop} leads to 
\begin{equation}
\int \frac{dk_0}{2\pi}\frac{d^dk}{(2\pi)^2} \frac{a_2 {\bf k}^2+m^2}{\tilde{k}^2-(a_2 {\bf k}^2+m^2)^2+\lambda_0+i\epsilon}-
\int \frac{dk_0}{2\pi}\frac{d^dk}{(2\pi)^2} \frac{a_2 {\bf k}^2+m^2}{\tilde{k}^2-(a_2 {\bf k}^2+m^2)^2+i\epsilon}=0,
\label{9.19}
\end{equation}
whereas the sum of the diagrams in the bottom of Fig. \ref{oneloop} implies,
\begin{equation}
\int \frac{dk_0}{2\pi}\frac{d^dk}{(2\pi)^2} \frac{i}{\tilde{k}^2-(a_2 {\bf k}^2+m^2)^2+\lambda_0+i\epsilon}-\frac{1}{2\tilde{g}}=0,
\label{9.20}
\end{equation}
where $d=1,2$. The gap equation (\ref{9.19}) implies that $\lambda_0=0$ and then supersymmetry is not spontaneously broken. With this, the Eq. (\ref{9.20}) reduces to
\begin{equation}
\int \frac{dk_0}{2\pi}\frac{d^dk}{(2\pi)^2} \frac{i}{\tilde{k}^2-(a_2 {\bf k}^2+m^2)^2+i\epsilon}=\frac{1}{2\tilde{g}}.
\label{9.21}
\end{equation}
In order to compare with the constraint equation of the supersymmetric spherical model, we need to integrate over the component $k_0$, which gives
\begin{equation}
\int \frac{d^dk}{(2\pi)^2} \frac{1}{\sqrt{a_1{\bf k}^2+(a_2 {\bf k}^2+m^2)^2}}=\frac{1}{ \tilde{g}}.
\label{9.22}
\end{equation}
There is one last step to compare this equation with the constraint equation (\ref{6}), which we rewrite here for convenience,
\begin{eqnarray}
1=\frac{\sqrt{g}}{2}\int \frac{d^d{q}}{(2\pi)^d}\frac{1}{\gamma+2U\sum_i \cos(q_i)}.
\label{9.23}
\end{eqnarray}
We need to take the continuum limit in this expression. To this, we reinsert the lattice spacing $a$ through ${q_i}\rightarrow a {q_i}$, and then take the limit $a\rightarrow 0$, so that the Brillouin zone $[-\frac{\pi}{a},\frac{\pi}{a}]$ extends to the infinity. Thus the integrals over momentum components become unlimited as in (\ref{9.22}). With this, the above equation becomes
\begin{eqnarray}
\frac{2}{\sqrt{g}a^{d-1}}=\int \frac{d^d{q}}{(2\pi)^d}\frac{1}{\left(\frac{\gamma-\gamma_c}{a}\right)+a|U|{\bf q}^2+O(a^3)},
\label{9.24}
\end{eqnarray}
recalling that $\gamma_c= 2d|U|$. Comparison of this expression with (\ref{9.22}) leads to the following identification of the parameters, 
\begin{eqnarray}
\frac{1}{\tilde{g}}~~&\Leftrightarrow&~~ \frac{2}{\sqrt{g}a^{d-1}},\nonumber\\
a_1~~&\Leftrightarrow&~~0\nonumber,\\
a_2~~&\Leftrightarrow&~~a|U|\nonumber,\\
m^2~~&\Leftrightarrow&~~\left(\frac{\gamma-\gamma_c}{a}\right),
\label{9.25}
\end{eqnarray}
completing thus the discussion of the equivalence of the two models.

As a final comment, we recall the computation of the correlation function with the presence of the term $\alpha \cos q$ in Sec. \ref{S5}. In that case we did not discuss how to further modify the theory in a way compatible with supersymmetry. This is automatic in the nonlinear sigma model, since it is associated with the term $a_1 {\bf k}^2$ in (\ref{9.22}). This comes from the bosonic propagator or equivalently from the bosonic quadratic part of the Lagrangian. The parameter $a_1$ is also included in the fermionic quadratic part of the Lagrangian (\ref{9.11})  as well as in the supersymmetry transformations (\ref{9.10}). We see clearly that the contributions involving $a_1$ yield to a relativistic  structure when $a_2\rightarrow 0$. In this sense, the oscillating behavior appearing in the correlation function (\ref{Correlation_4}) for $\alpha\neq 0$ can be thought as due to a competition between Lorentz $(z=1)$ and Lifshitz $(z=2)$ scalings.

\section{Discussions and Conclusions}\label{S7}

We conclude the work with a brief summary of the main results along with additional comments. 
We  presented here an extensive analysis of the critical behavior of supersymmetric quantum spherical spins for the case of short-range interactions. Starting with the case of zero temperature, we found that the system undergoes a quantum phase transition without spontaneous supersymmetry breaking. In particular, for dimensions $2<d<4$, the critical behavior is nontrivial in the sense that it is not characterized by mean-field critical exponents. Of course, above the upper critical dimension we recover the mean-field results. 

In the case of finite temparature the supersymmetry is always broken by thermal effects. This allows for an additional saddle point solution with $\mu\neq 0$, which is not available in the case of zero temperature. In our analysis, we kept the parameter $\mu$ fixed at an arbitrary value compatible with the saddle point conditions. 
Hence, $\mu$ can be viewed as defining a one-parameter family of models which, interestingly, splits into two universality classes as we consider the solutions with $\mu=0$ and $\mu\neq 0$. For $\mu\neq 0$, the model exhibits nontrivial critical behavior for dimensions $2<d<4$, whereas that for $\mu=0$, nontrivial critical behavior occurs for $4<d<6$. 

Among the quantities studied, it interesting to further discuss the fermionic condensate $\langle \sum_{\bf{r}} \bar{\psi}_{\bf{r}} \psi_{\bf{r}}\rangle$, since it exhibits an unusual  temperature dependence. In principle, we would expect that any quantity involving the pairing of degrees of freedom would be destroyed by thermal fluctuations. 
However, our results show that the fermionic condensate actually increases with $T$. This is a consequence of the constraint structure of the model. 
To further appreciate this point, one may compute the fermionic condensate from the thermodynamic identity $\frac{\partial f}{\partial H_F}$, without taking into account that all the parameters are tied by the saddle point equations (\ref{25}) and (\ref{26}).  In doing this, we are in effect ignoring the constraints of the model. In this case, it is easy to see that the condensate does vanish for very high temperatures. Now, taking into account the constraint equations, this affects the dependence on the temperature of the fermionic condensate, producing the unusual behavior shown in (\ref{fc0}).

The  dependence of the fermionic condensate on $T$ is not very sensitive to the specific form of the interaction in the model. To understand this, we notice first that for $T<T_c$, Eq.~(\ref{35}) implies that all the temperature dependence comes from the factor $\beta$ in front of the integral, since $\gamma$ is fixed at $\gamma_c$, regardless the form of the interaction. 
Moreover, for $T>T_c$ the numerical analysis shows that the behavior of the condensate is dictated essentially by the limit of high temperatures, where also the interaction has little influence. Therefore, the temperature dependence is more closely related with the constraint structure of the model, and this is supported by the fact that the same behavior is observed in the model with mean-field interactions \cite{Lucas}.

Although the investigation of the one-dimensional correlation function in Sec. \ref{S5} shows that the model does not have competing interactions, it also indicates that with a suitable deformation of the theory we can in principle generates such competing orderings. Of course, care is needed to deform the theory in a way compatible with supersymmetry. While this is automatic in the field theoretical corresponding model, it is less obvious in terms of  quantum spherical spins. Nevertheless, it constitutes an interesting effect to be pursued in future investigations, mainly due to the potential to produce rich phase diagrams with modulated phases, what can be useful in the applicability of the model. 

Finally, the connection with the supersymmetric nonlinear sigma model extends the series of equivalences between variations of spherical models and the large $N$ limit of field theoretical models with short-range interactions. In addition to placing the supersymmetric quantum spherical spins in a broader context, this helps to alleviate the issues raised up by the (spatial) nonlocal nature of the constraints of the model. 

The main efforts henceforth are in an attempt to establish applications of the present studies in concrete physical systems. Given the rich phenomenology involved, we do expect that this can be achieved in several situations.


\section{Acknowledgments}

We acknowledge the financial support from the Brazilian funding agencies CAPES, CNPq, and Funda\c c\~ao Arauc\'aria. This work is dedicated to the memory of our colleague and friend Sergio Augusto Carias de Oliveira. 


\end{document}